\documentclass[letterpaper]{article}

\usepackage[T1]{fontenc}

\usepackage{geometry}
\usepackage{setspace}
\usepackage{tikz}
\usepackage{booktabs}
\usepackage{textcomp}
\usepackage[style = chem-acs, articletitle=true]{biblatex}
\usepackage{graphicx}
\usepackage{float}
\usepackage{subcaption}
\newfloat{scheme}{htbp}{los}
\floatname{scheme}{Scheme}
\floatname{chart}{Chart}
\newfloat{graph}{htbp}{loh}

\usepackage{chemformula} 
\usepackage[version = 4]{mhchem} 

\usepackage{authblk}
\author[1,2]{Michael R. Wagner}
\affil[1]{SimVantage, Inffeldgasse 13, Graz, 8010, Austria}
\affil[2]{IPPT, Graz University of Technology, Inffeldgasse 13, Graz, 8010, Austria}
\author[3]{Manuela Dubacher}
\affil[3]{Takeda Pharma Austria, Technologiestraße 5, Vienna, 1120, Austria}
\author[4]{Nikoletta Patsaki}
\author[1]{Philipp Eibl}
\author[4]{Peter Varun Dsouza}
\affil[4]{Research Center Pharamceutical Engineering GmbH, Inffeldgasse 13, Graz, 8010, Austria}
\author[5]{Michael Dekner}
\affil[5]{Independent Expert, Felbigergasse 119, Vienna, 1140, Austria}
\author[1,4]{Christian Witz}
\author[6]{Johan Remmelgas}
\affil[6]{AstraZeneca AB R\&D, Pepparedsleden 1, M\"olndal, 431 83, Sweden}
\author[4]{Stefan Reimann-Zitz}
\author[2,4]{Johannes Khinast*}

\title{Mixing of miscible liquids: Dimensionless scaling for intermediate-to-large density differences in a stirred tank}
\date{*Email: khinast@tugraz.at}

\begin{document}

\maketitle

\begin{abstract}
Mixing of miscible liquids is an essential process in multiple industrial settings, usually with the intent to homogenize the product. 
This seemingly simple process is in fact a complex hydrodynamic problem that has a direct impact on the product quality. 
In this study, numerical simulations of a stirred tank were performed with a 50/50 ratio of liquids and systematically varied the Reynolds and Richardson numbers. 
A positive correlation between the mixing time and the Richardson number was observed, as reported in the literature. 
The influence of the Reynolds number was not as pronounced and clear.  
Based on the Power, Froude and Richardson numbers, we were able to derive an exponential scaling for the dimensionless mixing time that collapsed all our data onto one master curve.
\end{abstract}

\section*{Keywords}

Dimensionless analysis, CFD, turbulent mixing, lattice Boltzmann method, stirred tank

\section{Introduction}

Mixing of fluids is one of the most abundant processes in nature, e.g., when fresh water mixes with sea water in a river delta or cold air hits warm air during a change in the weather. 
Also in industry, mixing of fluids is performed in numerous production processes.
Consequently, effective mixing and the optimization of the associated mixing system are of critical importance. 
Over the centuries, several mixing devices have been developed on various scales, ranging from micromixers (e.g., to produce lipid nanoparticles~\cite{shenAdvancedManufacturingNanoparticle2024,ripollOptimalSelfassemblyLipid2022,maekiUnderstandingFormationMechanism2017,hussainProductionMRNALipid2024}) to large industrial stirred tank reactors with volumes of several hundred cubic meters, e.g., for wastewater treatment ~\cite{gargouriApplicationContinuouslyStirred2011}.
Additionally, stirred tanks are widely used for cultivating cells in the production of pharmaceutical products~\cite{fangApplicationBioreactorTechnology2022}

The mixing of miscible liquids is a fundamental unit operation in process engineering, aiming to homogenize a mixture of two or more liquids. 
This process can occur naturally (e.g., through diffusion or buoyancy-driven convection) or be induced by agitation and convection (e.g., during mixing in the pharmaceutical, chemical, food and waste treatment industries). 
Inducing turbulence massively enhances the mixing efficiency due to the superimposed random fluctuations of the flow field. In comparison, laminar mixing relies on striation formation via stretching and folding and molecular diffusion, which is slow, compared to turbulent mixing.

In stirred vessels, mixing performance is commonly quantified by the mixing time, defined as the time required to reduce concentration inhomogeneities below a prescribed level. 
Experimentally, it is typically determined from tracer experiments as the time elapsed after tracer injection until the measured concentration at predefined probe locations remains within $\pm5\%$ or $\pm10\%$ of its final value. 
Tracer experiments are widely used because they allow the characterization of mixing in a single homogeneous liquid phase~\cite{kasat_mixing_2004, bouwmans_blending_1997, mf_scalar_1997, distelhoff_scalar_2001, cabaretNewTurbineImpellers2008, moo-youngBlendingEfficienciesImpellers1972}. 

In simulations, mixing is usually quantified by the coefficient of variation (CoV), defined as the relative standard deviation of the concentration field. 
The temporal decay of probe concentration fluctuations in experiments is conceptually equivalent to the decay of the CoV, as both describe the reduction of relative concentration inhomogeneities.

A small amount of tracer leads to an injection that is effectively instantaneous in time and localized in space, closely resembling a Dirac-type impulse as the initial location. 
While mixing of a small quantify of fluid with a high-volume bulk is done occasionally, mixing of fluids with comparable amounts is more common in industrial systems. 
In those systems the liquids may differ in density.

Differences in density can induce buoyancy-driven stratification, i.e., the formation of vertically layered density regions in which gravitational forces counteract convective transport and suppress vertical mixing. 
Consequently, achieving a homogeneous mixture may necessitate increased power input at the impeller or longer mixing times~\cite{yuMixingStratifiedMiscible2018}. 

A considerable body of research has been dedicated to mixing in baffled stirred tanks~\cite{kramersComparativeStudyRate1953, mf_scalar_1997, distelhoff_scalar_2001, mayorgaReconstruction3DHydrodynamics2022}. 
Baffles allow higher stirrer velocities without the formation of a vortex. 
Baffles also create strongly turbulent flows above a certain critical Reynolds number, and thus, lead to a shorter mixing time~\cite{etchells_blending_2023}. 
Despite extensive studies of the hydrodynamics in stirred tanks using experimental methods and computational fluid dynamics (CFD) simulations~\cite{wuLaserDopplerMeasurementsTurbulentflow1989, sommerSolidliquidFlowStirred2021}, deriving generally valid correlations for the mixing time remains challenging, particularly when buoyancy effects are present.

To establish a reference, we first consider single-phase tracer mixing in a homogeneous Newtonian liquid under turbulent conditions. 
Ruszkowski~\cite{ruszkowski1994rotational} proposed a correlation for the dimensionless mixing time $t_m^{\ast}$, which is essentially the number of revolutions of the stirrer~\cite{hartmannMixingTimesTurbulent2006}:
\begin{equation}\label{eq:dimlessMT}
    t_m^{\ast} = N\cdot t_m,
\end{equation}
\begin{equation}\label{eq:Rusz}
    N\cdot t_m = 5.3\cdot N_p^{1/3}\left(\frac{T}{D}\right)^2,
\end{equation}
where $N$ is the impeller rotational speed, $t_m$ is the mixing time, $T$ and $D$ are the tank and impeller diameters, respectively, and $N_p$ is the power number,
\begin{equation}\label{eq:powerN}
    N_p = \frac{P}{\rho\cdot N^3 D^5},
\end{equation}
with $P$ denoting the power input and $\rho$ the density of the single liquid. 
The mixing time defined by Ruszkowski corresponds to the 95\% tracer mixing time, i.e., the time required for the tracer concentration to reach and remain within $\pm5\%$ of its final homogeneous value throughout the vessel. 
This correlation has been used and extensively cited in both numerical and experimental work~\cite{paulHandbookIndustrialMixing2010, nienowImpellerCirculationMixing1997,derksenBlendingMiscibleLiquids2011}. 
Subsequent experimental and CFD-based studies have corroborated the functional form of Eq.~(\ref{eq:Rusz}), yielding comparable proportionality constants for fully turbulent, single-phase blending in baffled stirred tanks~\cite{grenvilleBlendingMiscibleLiquids2003, zadghaffariMixingStudyDoubleRushton2009}.

In the fully turbulent regime, Eq.~(\ref{eq:Rusz}) implies that the dimensionless mixing time becomes independent of N but not the Reynolds number:
\begin{equation}\label{equ:re}
    Re = \frac{\rho\cdot N\cdot D^2}{\eta},
\end{equation}
where $\eta$ is the dynamic viscosity and $\rho$ is the density of the liquid. 
Consequently, further increases in impeller speed do not significantly reduce the dimensionless mixing time once fully developed turbulence is established. 

When density differences between two liquids induce stratification, buoyancy effects introduce an additional governing parameter~\cite{vandevusseMixingAgitationMiscible1955, ahmad1985mixing, rielly1988mixing, bouwmans_blending_1997}. 
CFD investigations confirm this trend and highlight the stabilizing effect of buoyancy on the interface between the layers~\cite{derksenBlendingMiscibleLiquids2011, derksenDirectSimulationsMixing2012} if the heavier liquid is at the bottom.

Yu et al.~\cite{yuMixingStratifiedMiscible2018} studied blending in a unbaffled stirred tank with a non-flat bottom and an off-center, inclined down-pumping axial impeller, differing from the classical baffled tank configurations typically considered, in the laminar-transitional regime, focusing on stratification effects and buoyancy. 
The results of their study suggested that buoyancy can be important for stratified systems (i.e., light fluid on top of the heavy fluid), especially when the lighter liquid has a volume fraction of more than a few percent. 
They found a good fit for the dimensionless mixing time using the relation
\begin{equation}\label{eq:yuMixing}
     N\cdot t_m = 30031\left(\frac{L_H}{T}\right)^{2.5}\left(\frac{D}{T}\right)^{-5.6} Re^{-1.3}Ri^{1.3},
\end{equation}
where $L_H$  denotes the liquid height in the vessel. 
In their work, the Richardson number is defined using the liquid height as the characteristic vertical length scale (i.e., $L_H=H$):
\begin{equation}\label{equ:ri}
    Ri = \frac{\Delta\rho}{\rho_h}\frac{g\cdot H}{(N\cdot D)^2},
\end{equation}
with $\Delta\rho$ being the density difference between the heavier ($\rho_h$) and the lighter liquid ($\rho_l$) and $g$ being the gravitational  acceleration.

While empirical correlations are well established for single-phase turbulent tracer mixing and case-specific scaling relations have been proposed for stratified systems, a generally valid framework that consistently accounts for gravitational, inertial, viscous and buoyancy effects across different geometries and flow regimes remain elusive.

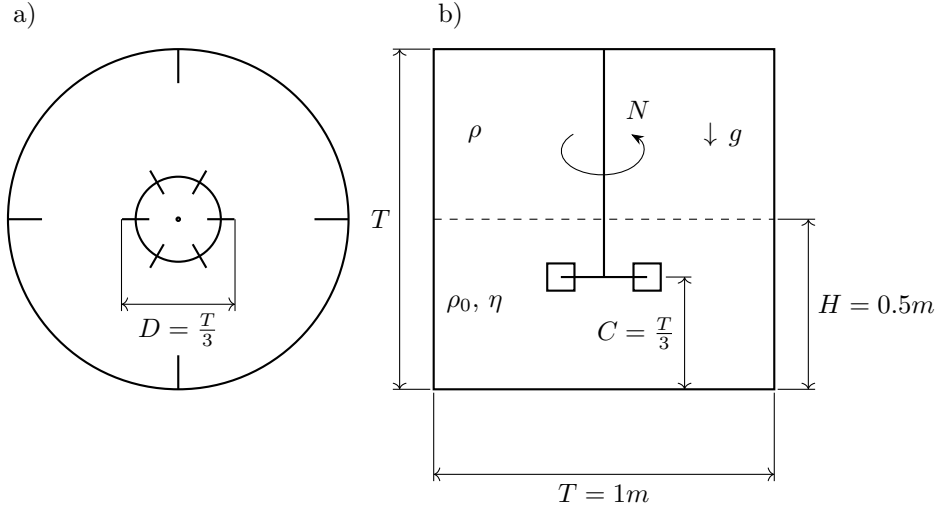
\begin{figure}
\centering
\begin{tikzpicture}[scale=4.5]
  \def\T{1.0}          
  \def\H{0.5}          
  \def\D{1/3}          
  \def\ybar{0.33}      
  \def\p{0.04}         
  \def\hshaft{0.26}    

  \draw[thick] (0,0) rectangle (\T,\T);
  \draw[dashed] (0,\H) -- (\T,\H);

  \draw[<->] (0,-0.25) -- (\T,-0.25) node[midway,below] {$T=1m$};
  \draw[<->] (1.1,0) -- (1.1,\H) node[midway,right] {$H=0.5m$};
  \draw[<->] (-0.1,0) -- (-0.1,\T) node[midway,left] {$T$};

  \draw[thick] (0.5-0.5*\D+\p,\ybar) -- (0.5+0.5*\D-\p,\ybar);

  \foreach \x in {0.5-0.5*\D+\p, 0.5+0.5*\D-\p} {
    \draw[thick] (\x-\p,\ybar-\p) rectangle (\x+\p,\ybar+\p);
  }

  \draw[thick] (0.5,\ybar) -- (0.5,\T);

  \draw[<->] (-0.75-0.5*\D,0.25) -- (-0.75+0.5*\D,0.25)
    node[midway,below] {$D=\tfrac{T}{3}$};
    
     \draw[<->] (0.57+0.5*\D,0.33) -- (0.57+0.5*\D,0)
    node[midway,left] {$C=\tfrac{T}{3}$};

  \node at (0.12,0.75) {$\rho$};
  \node at (0.12,0.25) {$\rho_0$, $ \eta$};
  \node at (0.85,0.75) {$\downarrow\, g$};
  
  \draw (0.515+0.5*\D,0.33) -- (0.59+0.5*\D,0.33);
  \draw (0,-0.01) -- (0,-0.27);
  \draw (\T,-0.01) -- (\T,-0.27);
  \draw (-0.01,0) -- (-0.12,0);
  \draw (-0.01,\T) -- (-0.12,\T);

  \draw (\T+0.01,0.5) -- (\T+0.12,0.5);
  \draw (\T+0.01,0) --(\T+0.12,0);
  \draw (-0.75-0.5*\D,0.5) -- (-0.75-0.5*\D,0.23);
  \draw (-0.75+0.5*\D,0.5) -- (-0.75+0.5*\D,0.23);

\def\ax{0.12}   
\def\ay{0.07}   

\draw[->,>=Stealth] (0.29+\ax,0.75)
  arc[start angle=-225, end angle=45, x radius=\ax, y radius=\ay];
\node at (0.6,0.82) {$N$};

\draw [thick] (-0.75,0.5) circle [radius=0.5];
\draw [thick] (-1.25,0.5)--(-1.15,0.5);
\draw [thick] (-0.25,0.5)--(-0.35,0.5);
\draw [thick] (-0.75,0)--(-0.75,0.1);
\draw [thick] (-0.75,0.9)--(-0.75,1);

\draw [thick] (-0.75,0.5) circle [radius=0.125];
\draw [thick] (-0.75,0.5) circle [radius=0.005];

\coordinate (C) at (-0.75,0.5);

\def\rinner{0.33/2} 
\def\router{0.33/2-0.08}  

\foreach \i in {0,60,120,180,240,300}{
    \draw[thick] (C) ++(\i:\rinner) -- ++(\i:{\router-\rinner});
}

  \node at (-1.2,1.1) {a)};
  \node at (0.05,1.1) {b)};
\end{tikzpicture}
\caption{Schematic representation of the stirred tank system with characteristic dimensions $T$, $H$, and $D$, used for the definition of Reynolds and Richardson numbers. 
The total liquid height equals the tank diameter $T$. 
The lighter liquid with density $\rho_l$ and viscosity $\eta$ is initially located above the interface at height $H$, while the heavier liquid with density $\rho_h$ and dynamic viscosity $\eta$ occupies the region below $H$. 
The impeller of diameter $D$ rotates counter clockwise at a rotational speed $N$. 
The impeller blade height is $D/5$ and the blade length is $D/4$. 
Four radial baffles of width $T/10$ are mounted flush with the tank wall, i.e., no gap between the baffles and the wall.}
\label{fig:system}
\end{figure}

In this study, we focused on a two-fluid system in a stratified state, i.e., a light liquid ($\rho_l$) placed on top of a heavy liquid ($\rho_h$) with $\rho_l=\rho_h-\Delta\rho$. 
The density difference $\Delta\rho$ is considered as a parameter, and no specific chemical identity is assumed. 
Both fluids are assumed to have identical dynamic viscosities. 
The basic schematic setup shown in Fig.~\ref{fig:system} was similar to the one in ref.~\cite{deglonCFDModellingStirred2006}.

\section{Method}
To establish mixing times for the system depicted in Fig.~\ref{fig:system}, we applied CFD to discretize and approximate the Navier-Stokes equation~\cite{Navier, Stokes} which for incompressible fluids reads
\begin{equation}\label{eq:NavierStokes}
    \partial_t\mathbf{u} + (\mathbf{u}\cdot\nabla)\mathbf{u} - \eta\Delta\mathbf{u} = -\frac{1}{\rho}\nabla p + \frac{1}{\rho}\mathbf{F},
\end{equation}
\begin{equation}\label{eq:conti}
    \nabla\cdot\mathbf{u} = 0,
\end{equation}
where $\mathbf{u}$ is the flow or velocity (vector-) field, $\nu$ is the kinematic viscosity, $\rho$ is the density, $p$ is the pressure and $\mathbf{F}/\rho$ is an arbitrary body-force acceleration term, e.g., the gravitational acceleration $(0,0,-g)$. 

\subsection{Lattice Boltzmann method}\label{sec:lbm}
\begin{figure}
\centering
\includegraphics[width=0.8\linewidth]{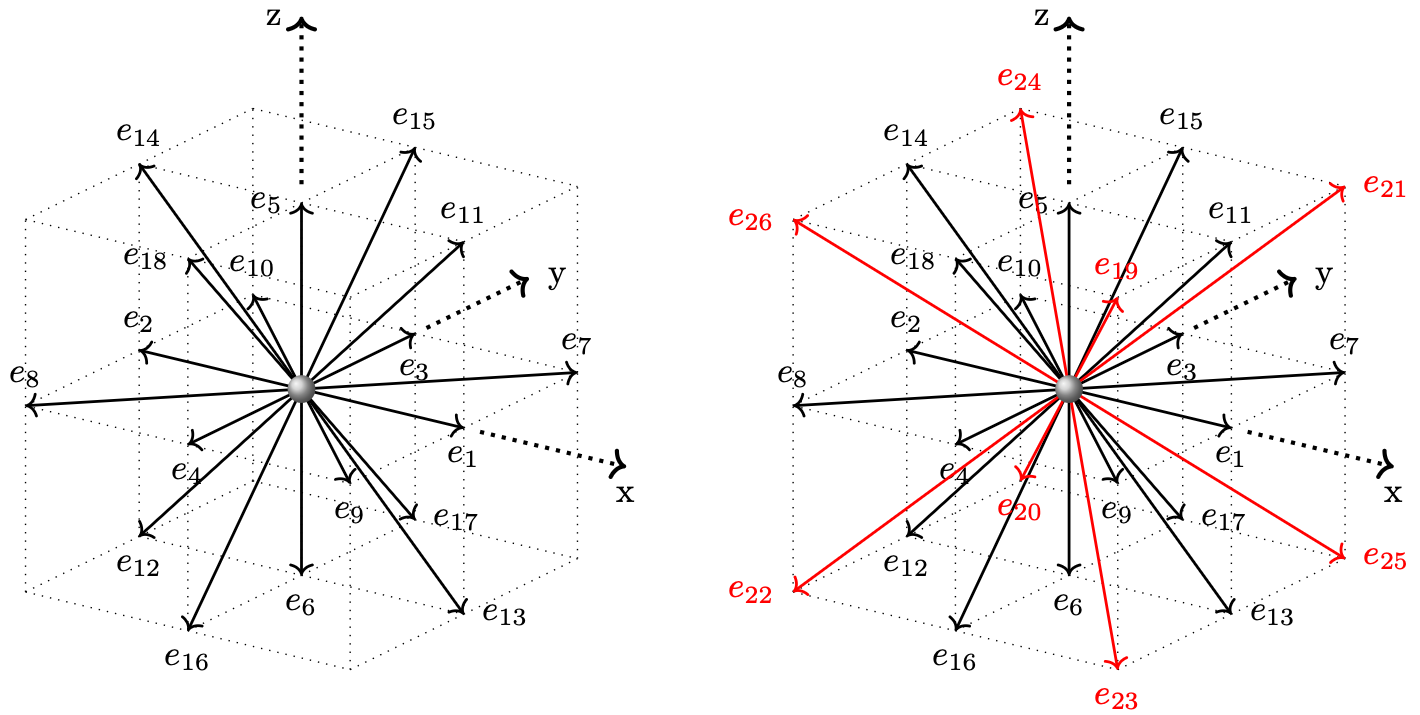}
\caption{LBM Scheme: D3Q19 left, D3Q27 right}
\label{fig:lbm_scheme}
\end{figure}
The lattice Boltzmann method (LBM) is one of many CFD approaches~\cite{krugerLatticeBoltzmannMethod2017a, succiLatticeBoltzmannEquation2001,heTheoryLatticeBoltzmann1997}. 
In contrast to many other methods such as the finite volume method (FVM) the LBM is derived from the Boltzmann equation. 
It uses $q$ populations $f_i(i = 0 , ... , q-1)$ which can be regarded as mesoscopic particle distributions, and which are allowed to stream and collide  
The streaming occurs along a regular lattice in discrete time steps, with the most common lattices are $D2Q9$, $D3Q19$ and $D3Q27$ (the two later ones depicted in Fig.~\ref{fig:lbm_scheme}). 
One of the surprising strengths of this method is that it only requires algebraic manipulations of the populations, which makes the method a perfect candidate for numerical highly parallelized implementation~\cite{krugerLatticeBoltzmannMethod2017a}.

Here the D3Q19 scheme as shown in Fig.~\ref{fig:lbm_scheme} (left) and solve the LBM master equation,
\begin{equation}\label{eq:lbm_master}
    f_i(\mathbf{x}+\mathbf{c}_i\Delta t,t+\Delta t) = f_i(\mathbf{x},t) + \Delta t \Omega_i + \Delta t F_i(\mathbf{x},t), 
\end{equation}
with $\mathbf{x}$ being the position vector, $\mathbf{c}_i$ being the $q$ discrete velocities given by 
\begin{equation}\label{eq:speeds}
\mathbf{c}_i  =
\left\{
\begin{array}{ll}
(0,0,0) & i = 0 \\
(\pm 1, 0, 0), (0, \pm 1, 0), (0, 0, \pm 1) &  i=1-6 \\
(\pm 1, \pm 1, 0), (\pm 1, 0, \pm 1), (0, \pm 1, \pm 1) & i=7-18
\end{array}
\right.,
\end{equation}
and $\Omega_i$ the collision operator 
\begin{equation}\label{eq:BGK}
    \Omega_i = -\frac{1}{\tau}(f_i(\mathbf{x},t) - f_i^{eq}(\mathbf{x},t)).
\end{equation}
In this work the single relaxation time $\tau$ (SRT) operator introduced by Bhatnagar, Gross and Krook (BGK)~\cite{bhatnagarModelCollisionProcesses1954} is used that redistributes the populations $f_i(\mathbf{x},t)$ towards the equilibrium.
The equilibria are derived as an expansion of the Maxwell-Boltzmann distribution up to second order:
\begin{equation}\label{eq:equilibria}
    f_i^{eq}(\mathbf{x},t) = w_i\rho\left[1 +\frac{\mathbf{c}_i\cdot\mathbf{u}}{c_s^2} + \frac{(\mathbf{c}_i\cdot\mathbf{u})^2}{c_s^4} - \frac{\mathbf{u}\cdot\mathbf{u}}{2c_s^2}\right],
\end{equation}
where $c_s = (1/\sqrt{3})\Delta x / \Delta t$ is the speed of sound and the weights $w_i$ are
\begin{equation}\label{eq:weights}
w_i  =
\left\{
\begin{array}{ll}
1/3 & i = 0 \\
1/16 &  i=1-6 \\
1/32 & i=7-18
\end{array}
\right..
\end{equation}
The force term in Eq.~(\ref{eq:lbm_master}) can be calculated according to~\cite{krugerLatticeBoltzmannMethod2017a, haussmannLargeeddySimulationCoupled2019, guo_discrete_2002} as:
\begin{equation}\label{eq:force}
    F_i(\mathbf{x},t) = \left(1 - \frac{1}{2\tau}\right)w_i\left[\frac{\mathbf{c}_i - \mathbf{u}}{c_s^2} +\frac{\mathbf{c}_i\cdot \mathbf{u}}{c_s^4}\mathbf{c}_i\right]\cdot\mathbf{g},
\end{equation}
with $\mathbf{g}$ being some external body acceleration.
Lastly the hydrodynamic quantities can be computed as discrete moments of $f_i$
\begin{align}
    \rho &= \sum_i^{q-1} f_i,\label{eq:LBMdensity}\\
    \mathbf{u} &= \frac{1}{\rho}\sum_i^{q-1} \mathbf{c}_if_i,\label{eq:LBMmomentum}\\
    \Pi_{\alpha\beta} &= \frac{1}{\rho}\sum_i^{q-1} \mathbf{c}_{i\alpha}\mathbf{c}_{i\beta}f_i.\label{eq:LBMtensor}
\end{align}

As some simulations are in a highly turbulent system ($Re > 10^5$), we model sub-grid turbulence using a the Smagorinsky-Lilly large eddy simulation (S-L LES) model~\cite{smagorinskyGENERALCIRCULATIONEXPERIMENTS1963, yuDNSDecayingIsotropic2005}. 
The S-L LES model introduces an effective or eddy viscosity $\eta_t$ which can be calculated based on the momentum flux. 
Further details on the specific implementation can be found in ref.~\cite{witzLocalGasHoldup2016}.

To describe the convective mixing of a miscible component, i.e., the lighter liquid phase ($\rho_l$), we adopted the mass–flux-based scalar transport model proposed by Osmanlic et al.~\cite{osmanlic_lattice_2016}. 
The scalar field $\phi$ is transport proportionally to the net mass flux of the underlying lattice Boltzmann fluid. 
In the present work, $\phi$ is additionally coupled to the momentum equation through a concentration dependent gravitational body force. 
This force is defined as the difference of the distribution functions:
\begin{equation}
    \Delta f_i = f_{\bar{i}}(\mathbf{x}+\mathbf{c}_i,t) - f_i(\mathbf{x},t),
\end{equation}
where $\bar{i}$ denotes the lattice direction opposite to $i$. The sign of $\Delta f_i$ determines whether $\phi$ is transported away from or towards the lattice node located at position~$\mathbf{x}$. 
The update of $\phi$ stored at a lattice node is then given by:
\begin{equation}
\Delta \phi_i
=
\sum_k
\begin{cases}
\dfrac{\Delta f_k}{\rho(\mathbf{x}+\mathbf{c}_k\Delta t,t)}\, \phi_i(\mathbf{x}+\mathbf{c}_k\Delta t,t),
& \Delta f_i \ge 0, \\[2.5ex]
\dfrac{\Delta f_k}{\rho(\mathbf{x},t)}\, \phi_i(\mathbf{x},t),
& \Delta f_i < 0 ,
\end{cases}
\end{equation}
where $\phi(\mathbf{x},t)$ denotes a concentration located at the lattice node at position $\mathbf{x}$ and time $t$.
All lattice directions are evaluated, resulting in a multidirectional advection scheme that is fully consistent with the discrete mass transport inherent to the LBM. 

In industrial stirred tanks diffusive transport plays only a minor role in mixing and can be neglected. 
The only diffusive effects that remain are of numerical origin. 
They have been analysed in detail by Küng et al.~\cite{kung_comparison_nodate}. 
Since density difference arising from local variations in scalar concentration are consistently accounted for, changes in the concentration of the lighter liquid that affect the local fluid density are fully reflected in the resulting body forces. 
Assuming a unit volume $V=1$ in lattice units, the body force is given by
\begin{equation}
    \mathbf{F}_b = (\rho_l - \rho_h)\mathbf{g}.
\end{equation}

As can be seen in Fig.~\ref{fig:system} the three types of boundary conditions for the geometry are included: the solid vessel walls, the impeller and the free surface. 
The solid vessel walls were modelled as a no-slip boundary condition using a simple bounce-back scheme~\cite{zhangGeneralBouncebackScheme2012}. 
At the free surface we applied a zero gradient or free slip boundary condition. 
The last remaining boundary was the solid surface of the stirrer, for which the modified bounce-back method~\cite{laddNumericalSimulationsParticulate1994} was used:
\begin{equation}\label{eq:modbb}
    f_{i}^{\star}(\mathbf{x},t+\Delta t) = f_{i}(\mathbf{x},t) - 2w_i\rho\frac{\mathbf{u}_b\cdot\mathbf{c}_i}{c_s^2}, 
\end{equation}
with $\mathbf{u}_b$ being the velocity of the solid.

\subsection{Parameters and simulation space}\label{sec:simulationsPerformed}
\begin{figure}
  \centering
  \begin{subfigure}{0.25\textwidth}
    \centering
    \includegraphics[width=\linewidth]{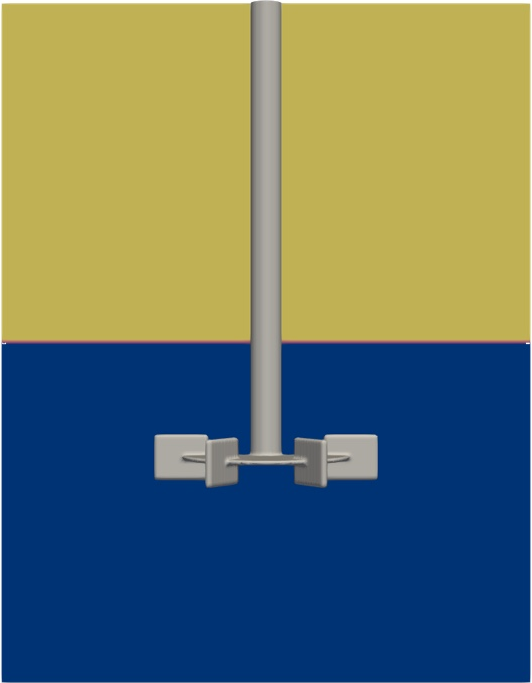}
    \caption{t = 0 s}
    \label{fig:mixing_0s}
  \end{subfigure}
  \begin{subfigure}{0.25\textwidth}
    \centering
    \includegraphics[width=\linewidth]{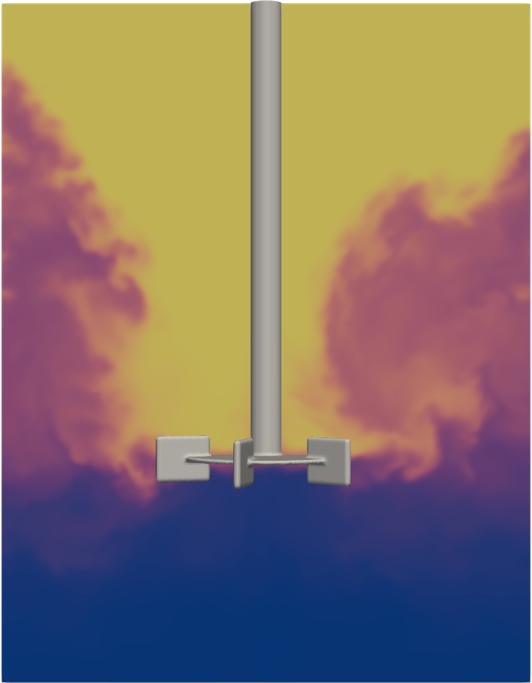}
    \caption{t = 30 s}
    \label{fig:mixing_30s}
  \end{subfigure}
  \begin{subfigure}{0.25\textwidth}
    \centering
    \includegraphics[width=\linewidth]{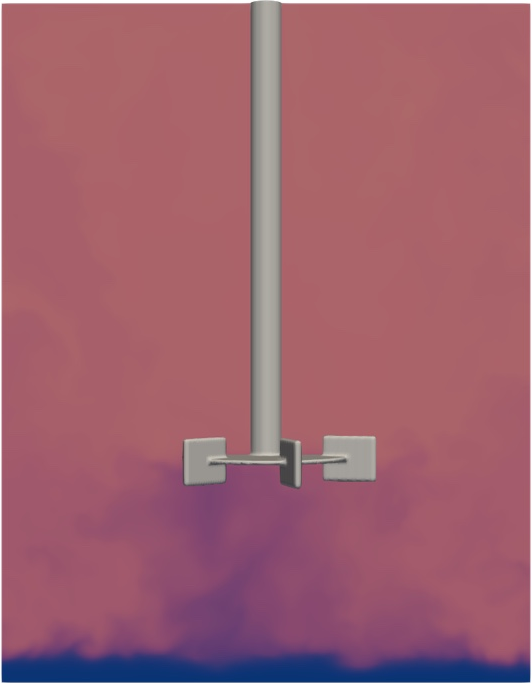}
    \caption{t = 50 s}
    \label{fig:mixing_50s}
  \end{subfigure}
  \begin{subfigure}{0.15\textwidth}
    \centering
    \includegraphics[height=4.8cm]{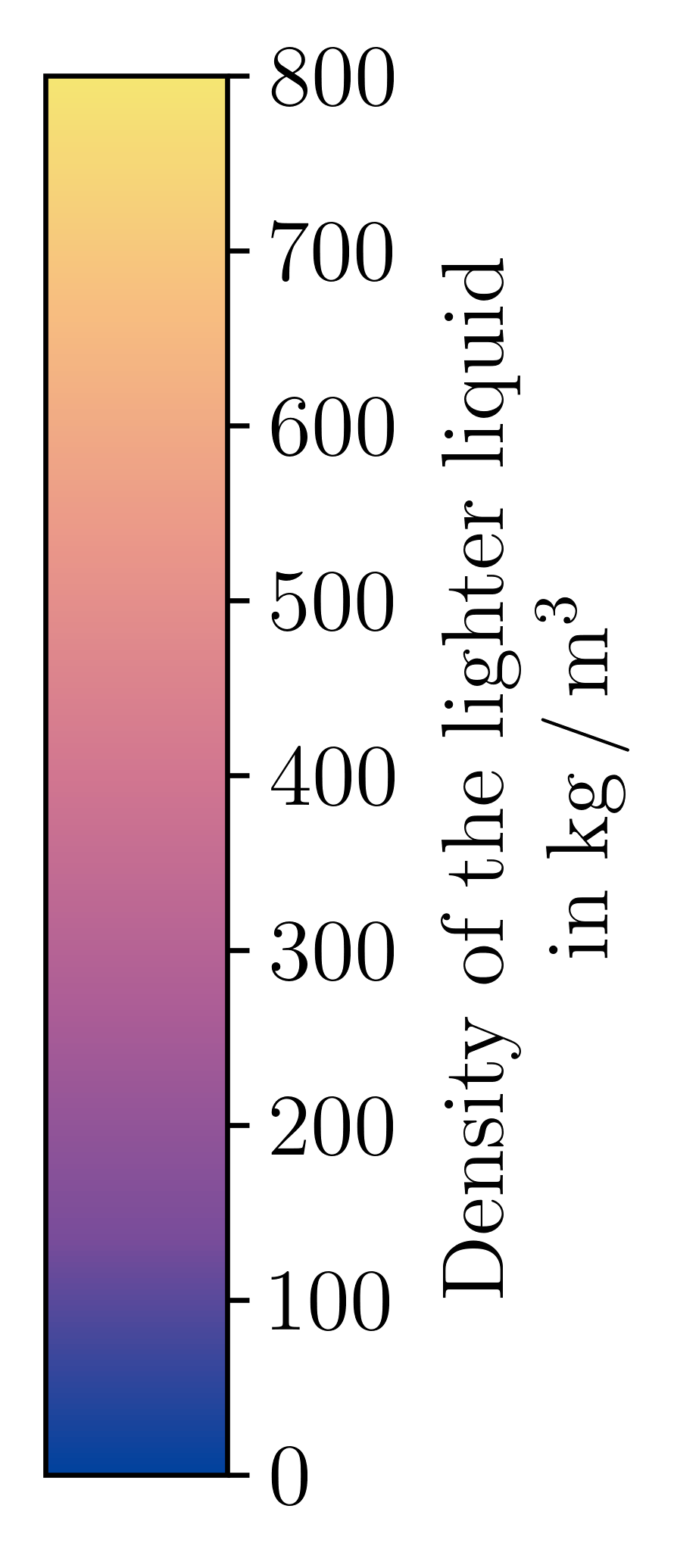}
    \vspace{8mm} 
  \end{subfigure}
  \caption{Temporal evolution of substrate concentration during mixing.
  The impeller induces progressive homogenisation of the initially stratified concentration field.}
  \label{fig:mixing_snapshots}
\end{figure}
\begin{table}
\centering
\caption{Substrate density in $kg/m^3$ for $Ri=0$ to $Ri=14$ and $Re=50000$ to $Re=125000$}
\label{tab:dens_all}
\begin{tabular}{c|cccccccc}
\toprule
 Re/Ri & 0 & 2 & 4 & 6 & 8 & 10 & 12 & 14 \\
\midrule
  50000  & 1000 & 990.83 & 981.65 & 972.48 & 963.30 & 954.13 & 944.95 & 935.78 \\
  62500  & 1000 & 985.67 & 971.33 & 957.00 & 942.66 & 928.33 & 913.99 & 899.66 \\
  75000  & 1000 & 979.36 & 958.72 & 938.07 & 917.43 & 896.79 & 876.15 & 855.50 \\
  87500  & 1000 & 971.90 & 943.81 & 915.71 & 887.61 & 859.52 & 831.42 & 803.33 \\
 100000  & 1000 & 963.30 & 926.61 & 889.91 & 853.21 & 816.51 & 779.82 & 743.12 \\
 112500  & 1000 & 953.56 & 907.11 & 860.67 & 814.22 & 767.78 & 721.33 &  -\\
 125000  & 1000 & 942.66 & 885.32 & 827.98 & 770.64 & 713.30 & - & - \\
\bottomrule
\end{tabular}
\end{table}
A set of simulations was performed with two parameters, i.e., stirrer speed $N$ and the density of the lighter phase $\rho_l$. 
The density of the heavy phase was kept constant at $1000 kg/m^3$. 
This affects the dimensionless numbers $Re$ and $Ri$ (see Eqs.~(\ref{equ:re}, \ref{equ:ri})). 
A broad range of parameters was chosen, with the Richardson number $Ri$ and the Reynolds number $Re$ ranging from 0 to 14 and from 50000 to 125000, respectively. 
$Ri = 0$ implies that both fluids have the same density, while larger Ri numbers reflect liquids with larger density differences. 
Since the density $\rho_l$ influences both $Re$ and $Ri$, the investigated cases listed in Table~\ref{tab:dens_all} are defined by expressing the density as a function of these dimensionless groups according to:
\begin{equation}
\label{equ:density}
    \rho = \rho_0 - \frac{Ri \cdot Re^2 \cdot \eta^2}{g \cdot \rho_0 \cdot H \cdot D^2},
\end{equation}
with $g$ being the gravitational acceleration $g = |\mathbf{g}| = 9.81\frac{m}{s^{2}}$.

\section{Results}
The numerical accuracy and reliability of the simulation were evaluated prior to analysing the mixing times. 
Like any other CFD method, LBM simulations are limited by the discretization scheme and the resolution. 
The numerical uncertainty was evaluated by computing the Grid Convergence Index (GCI) to ensure sufficient spatial resolution. 
Furthermore, a validation simulation of a different stirred tank system with probes to monitor concentrations was conducted to compare the mixing times to the experimental results reported in~\cite{fangApplicationBioreactorTechnology2022}. 
For completeness, the Appendix presents the simulated responses for four of the six installed probes, the results of the grid convergence study and an examination of the relationship between the normalized probe response regarding the concentration and the CoV.

Beyond this validation study a free surface model was additionally employed to analyze the case with the highest Reynolds number ($Re = 125000$) with respect to its possible free-surface dynamics. 
For this purpose, the LBM volume-of-fluid approach implemented in SimVantage~\textregistered~was used to assess whether significant surface-related effects might have been neglected~\cite{bognerCurvatureEstimationVolumeoffluid2016}. 
The results presented in the Appendix indicate that the free surface remains largely flat for these conditions. 
The mixing times obtained for all simulated cases are analyzed in dimensionless form defined in Eq.~\ref{eq:dimlessMT}, corresponding to the number of impeller revolutions required to reach $90\%$ homogeneity. 
A typical simulation result is shown in Fig.~\ref{fig:mixing_snapshots} at three-time instances with a color code indicating the density

\begin{figure}
    \centering
    \includegraphics[width=0.48\linewidth]{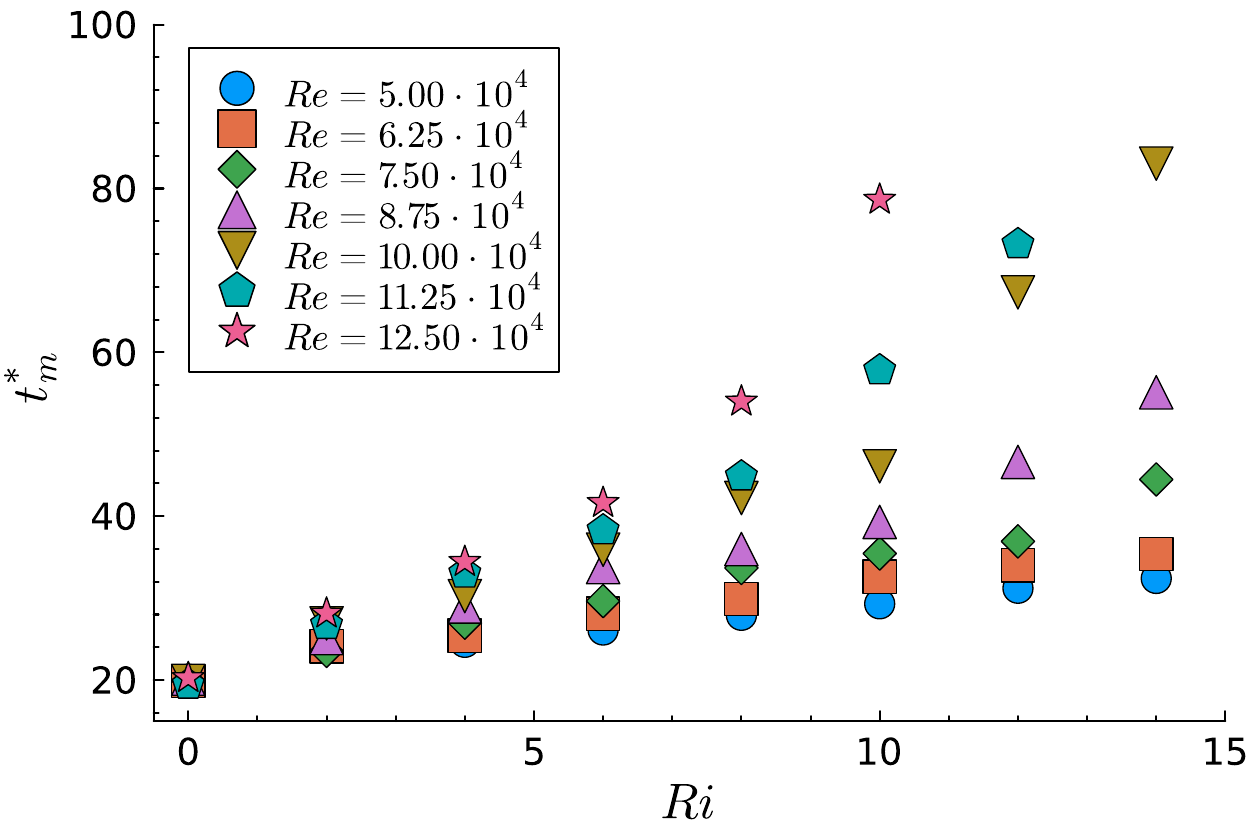}
    \includegraphics[width=0.48\linewidth]{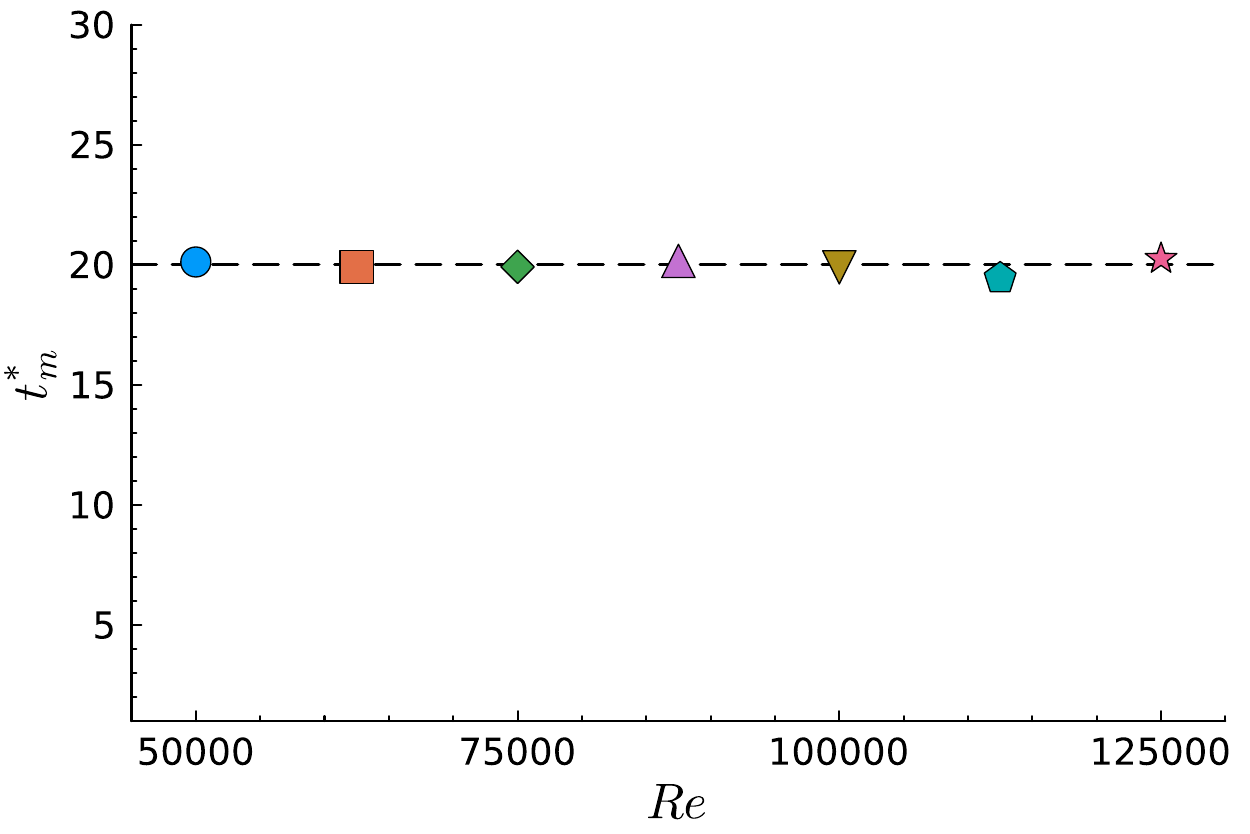}
    \caption{Left: Dimensionless mixing times $t_m^{\ast}$ as a function of Ri. 
    Different symbols depict different Re values. 
    Right: $t_m^{\ast}$ at Ri=0 for different Re values. 
    As a visual guide, a dashed line is shown at $t_m^{\ast}=20$.}
    \label{fig:tm_nd_Ri}
\end{figure}

\begin{table}
\centering
\caption{Dimensionless mixing time $t_m^{\ast}$ for $Ri=0$ to $Ri=14$ and $Re=50000$ to $Re=125000$.
Cases where $\rho < 700kg/m^3$ have been neglected.}
\label{tab:mix_time}
\begin{tabular}{c|cccccccc}
\toprule
 Re/Ri & 0 & 2 & 4 & 6 & 8 & 10 & 12 & 14 \\
\midrule
 50000  & 20.1 & 23.8 & 24.6 & 26.1 & 27.9 & 29.3 & 31.2 & 32.4 \\
 62500  & 19.9 & 24.1 & 25.4 & 28.1 & 29.9 & 32.6 & 34.0 & 35.4 \\
 75000  & 19.9 & 23.6 & 27.0 & 29.6 & 33.7 & 35.4 & 36.9 & 44.5 \\
 87500  & 20.2 & 25.1 & 29.0 & 33.8 & 36.0 & 39.3 & 46.6 & 55.1 \\
100000  & 19.9 & 26.9 & 30.2 & 35.9 & 42.2 & 46.1 & 67.3 & 83.0 \\
112500  & 19.4 & 26.7 & 33.0 & 38.3 & 44.9 & 57.8 & 73.2 & -- \\
125000  & 20.3 & 28.1 & 34.4 & 41.6 & 54.0 & 78.6 & --   & -- \\
\bottomrule
\end{tabular}
\end{table}
The resulting values are summarized in Table~\ref{tab:mix_time} and plotted in Fig.~\ref{fig:tm_nd_Ri} (left) as a function of the Richardson number for different Reynolds numbers. 

In the absence of buoyancy effects ($Ri=0$), all dimensionless mixing times collapse to a constant value of $t_m^{\ast}\approx 20$ independent of the Reynolds number in the investigated range $5\times10^4\leq Re \leq 1.25\times 10^5$. 
This value is consistent with literature data. 
Bouwmans et al.~\cite{bouwmans_blending_1997} report a dimensionless $95\%$ mixing time of $t_{m,95}^{\ast} = 26$. 
Assuming an exponential decay of concentration fluctuations during the final homogenization stage, the mixing time corresponding to different homogenization criteria scales as $t_{m, \epsilon} \propto \ln(1)/\epsilon$, where $\epsilon$ denotes the threshold value of the normalized concentration fluctuations, e.g., the CoV, used as the mixing criterion.
Consequently applying
\begin{equation}\label{eq:mixingtime9095}
    t_{m,95} = \frac{\ln(20)}{\ln(10)} \approx 1.3\cdot t_{m,90},
\end{equation}
to the present results, demonstrates good quantitative agreement once identical mixing criteria are applied. 
Such an invariance confirms that the flow is fully turbulent and that inertial mixing dominates.

\begin{figure}
    \centering
    \includegraphics[width=0.65\linewidth]{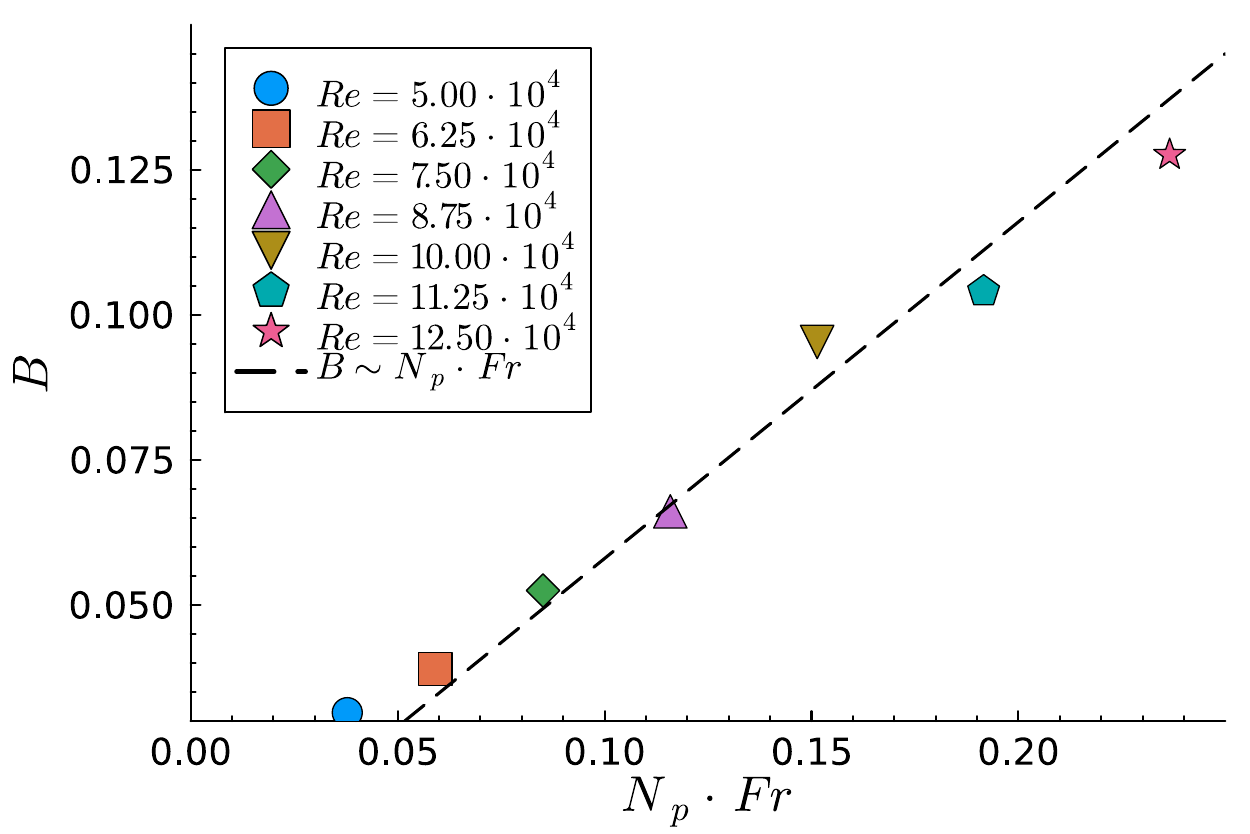}
    \caption{Plot of the best fit $B$ values as a function of $N_p$ $Fr$ for different $Re$ values. 
    The dashed line indicates Eq.~(\ref{eq:Bvalue}).}
    \label{fig:Bvaluefit}
\end{figure}

When buoyancy effects are introduced ($Ri>0$), the dimensionless mixing time increases systematically with increasing Richardson number. 
As shown in Fig. 4 and detailed in Table~\ref{tab:mix_time}, the data follows an exponential trend across all Reynolds numbers. 
The results can be represented by
\begin{equation}\label{eq:expModel}
    t_m^{\ast} = A\cdot e^{B\cdot Ri},
\end{equation}
with $A=20$, corresponding the the inertial baseline identified above.
Analysis of the parameter $B$ reveals a proportionality to the product of power number $N_p$ and Froude number
\begin{equation}\label{eq:fr}
    Fr = \frac{N^2\cdot D}{g},
\end{equation}
yielding the correlation,
\begin{equation}\label{eq:Bvalue}
    B = 0.58\cdot N_p\cdot Fr,
\end{equation}
as shown in Fig.~\ref{fig:Bvaluefit}.

\begin{figure}
    \centering
    \includegraphics[width=0.65\linewidth]{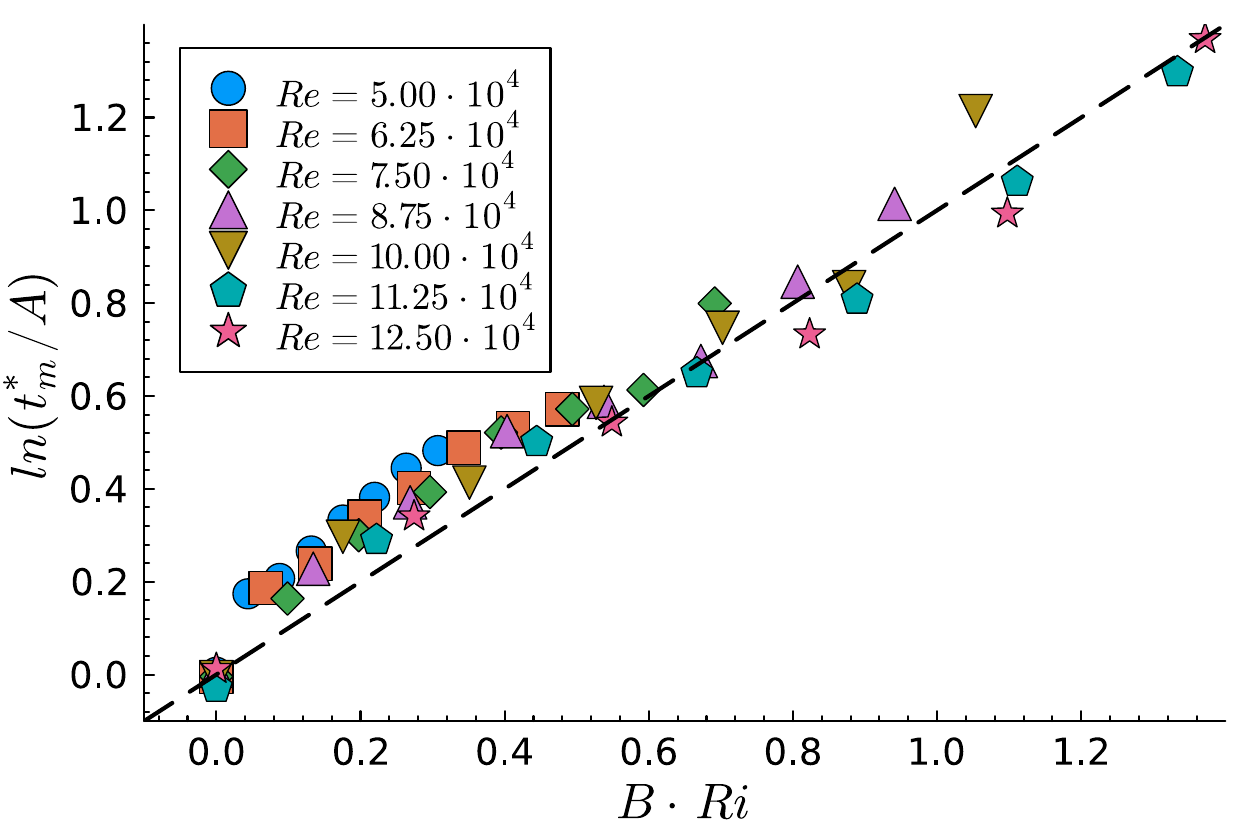}
    \caption{$t_m^{\ast}$ data plot against our model Eq.~(\ref{eq:masterus} using $A=20$ and $B=0.58\cdot N_p\cdot Fr$.}
    \label{fig:datacollapse}
\end{figure}

In the Reynolds-number range considered, the power number $N_p$ is approximately constant ($N_p=5.5$) since the system operates in the fully turbulent regime, as can be seen in the appendix. 
Using this relation leads to the final correlation
\begin{equation}\label{eq:masterus}
    t_m^{\ast} = 20\cdot e^{0.58\cdot N_p \cdot Fr \cdot Ri},
\end{equation}
which collapses the entire investigated parameter space onto a single master curve shown in Fig.~\ref{fig:datacollapse}.
The correlation captures the combined influence of inertial stirring and buoyancy stabilization and provides a compact representation of turbulent mixing in initially density-stratified systems. 
Figure 7 shows the agreement of our model for $t_m^{\ast}$ for the individual Reynolds numbers over the Richardson number.

With this scaling relation established, the behavior of the dimensional mixing times shown in Fig. 8 can be interpreted mechanistically. 
The graph in Fig. 8 shows the mixing time dependency on $Re$ at constant values of $Ri$. For small to intermediate Richardson numbers ($Ri<8$), the mixing time decreases monotonically with increasing Reynolds number which agrees with literature~\cite{wesselinghMixingLiquidsCylindrical1975, kasat_mixing_2004}. 
For larger Richardson numbers ($Ri\ge 8$), a non-monotonic dependence emerges, with a minimum mixing time at intermediate Reynolds number. 
\begin{figure}
    \centering
    \includegraphics[width=0.65\linewidth]{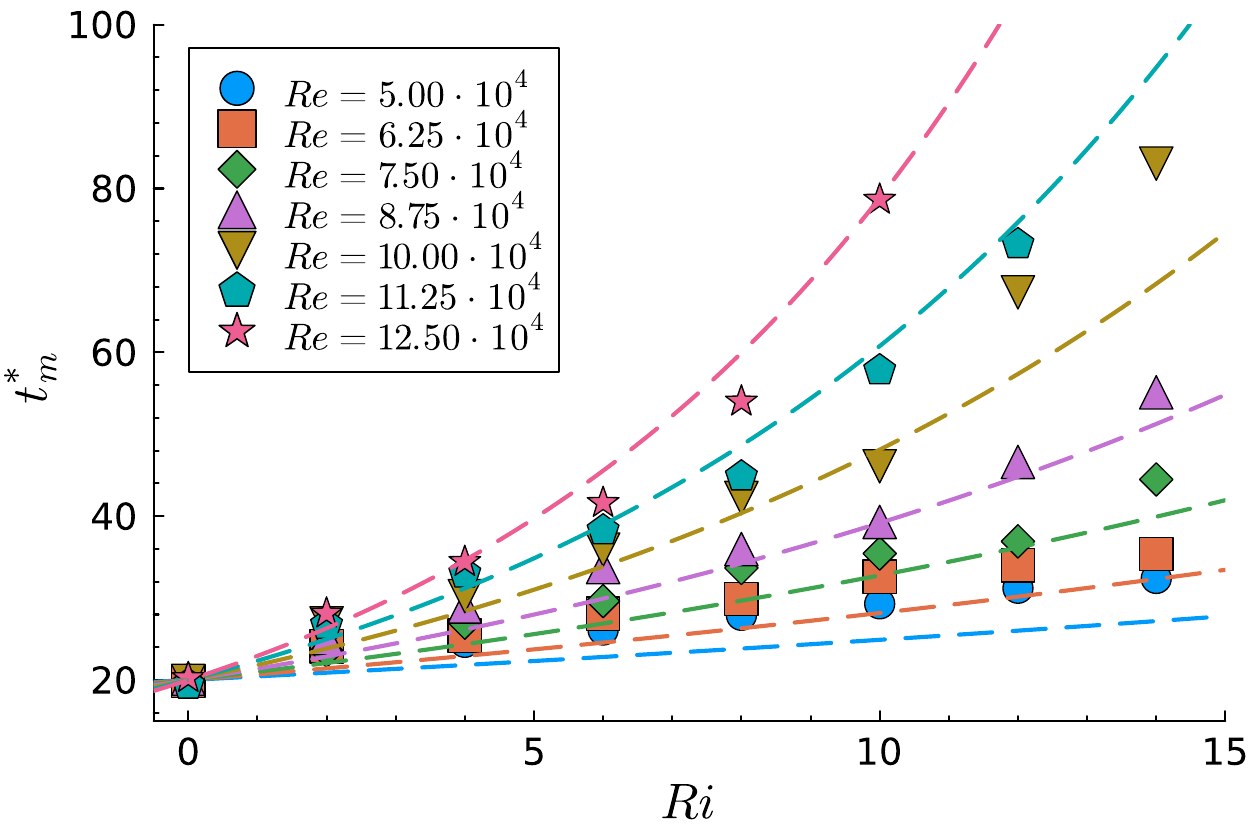}
    \caption{Dimensionless mixing time $t_m^{\ast}$ over $Ri$. 
    Our data is depicted with symbols, the full lines are our model.}
    \label{fig:placeholder}
\end{figure}

This behavior can be explained by noting that, according to Eqs.~(\ref{equ:re}), (\ref{equ:ri}) and (\ref{equ:density}), the Reynolds number, the Richardson number and the density difference $\Delta\rho$ are not independent. 
For a constant $Ri$ series, increasing $Re$ requires a simultaneous increase in $\Delta \rho$ (see Tab.~\ref{tab:dens_all}), leading to stronger buoyancy-induced stratification. 
While higher $Re$ enhances inertial stirring and turbulence production, stratification suppresses vertical transport. 
The interplay between shear production and buoyancy stabilization, governed by $Ri$~\cite{fernandoTurbulentMixingStratified1991}, can therefore result in a non-monotonic mixing response with a minimum at intermediate Reynolds numbers.

This behavior is directly reflected in the proposed correlation. 
Dividing Eq.~(\ref{eq:masterus}) by the impeller speed gives the dimensional mixing time as
\begin{equation}\label{eq:mixingTimeDim}
    t_m = \frac{20}{N}e^{0.58\frac{N_p\cdot H}{\rho_h D}\Delta\rho}.
\end{equation}
Notably, for $Re < 1.125\times 10^5$ the corresponding impeller speeds are below $N=1s^{-1}$. 
The $1/N$ term represents faster inertial mixing at higher stirrer speeds, while the exponential term captures the increasing buoyancy effect with larger density differences. 
In a constant-$Ri$ series, increasing $Re$ implies both higher $N$ and larger $\Delta\rho$:
\begin{equation}
    t_m \propto \frac{1}{N}e^{\alpha\Delta\rho}.
\end{equation}

The correlation thus captures the competition between enhanced turbulence and strengthened stratification, leading to a minimum mixing time at intermediate Reynolds numbers. 
\begin{figure}
    \centering
    \includegraphics[width=0.65\linewidth]{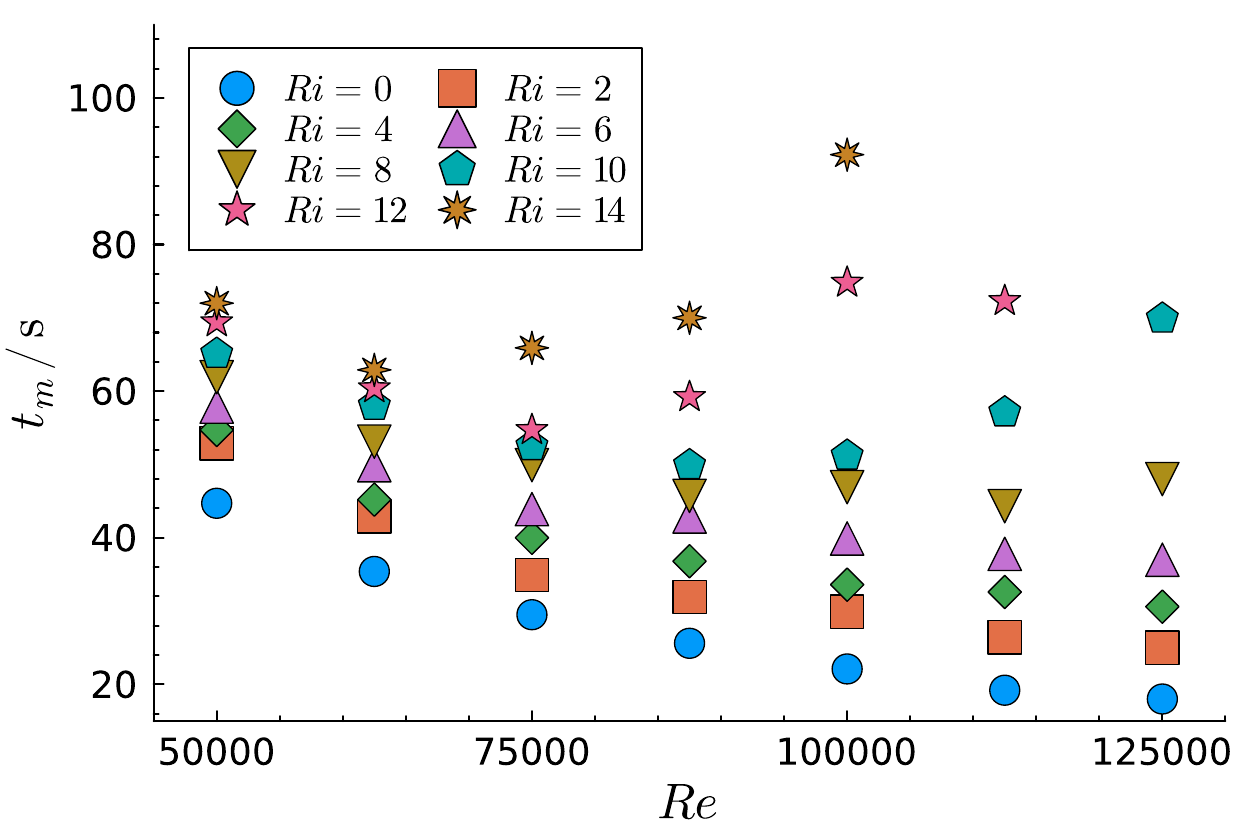}
    \caption{Mixing time $t_m$ as a function of $Re$ with different symbols for different constant $Ri$.}
    \label{fig:dimMixingTime}
\end{figure}
In conclusion, Fig.~\ref{fig:dimMixingTime} indicates that the $Ri$ number (and thus buoyancy) have a significant effect on $t_m$, as reported in the literature~\cite{derksenBlendingMiscibleLiquids2011, derksenDirectSimulationsMixing2012}. 
The corresponding velocity magnitude fields for three $Ri$ values at identical Reynolds number are shown in Fig.~\ref{fig:velocity_snapshots}. 
While the maximal velocity is overall independent of Ri since $Re\sim N\sim |u|$, the flow structures are different in the three cases. 
Especially in the case of $Ri=12$ depicted in Fig.~\ref{fig:velocity_snapshots}(c), we observed that the lighter phase was ``sealed off'' from the power input of the stirrer and that vertical structures only appear in the denser liquid.
This behaviour can be attributed to the strong density difference at high $Ri$, which suppresses vertical momentum transport and limits interfacial entrainment, consistent with the buoyancy-induced stabilization of turbulence in stratified flows~\cite{fernandoTurbulentMixingStratified1991}.
\begin{figure}
  \centering
  \begin{subfigure}{0.25\textwidth}
    \centering
    \includegraphics[width=\linewidth]{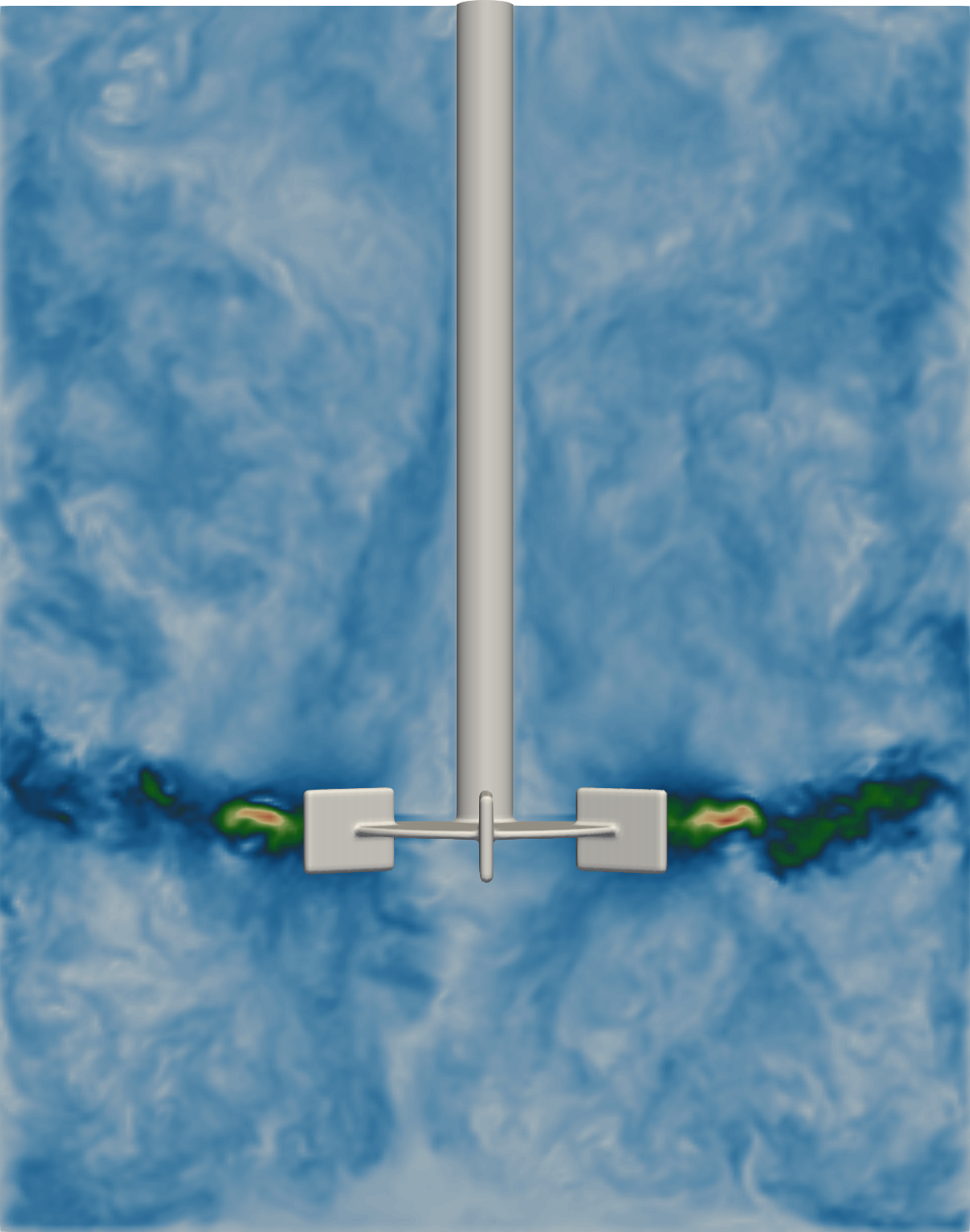}
    \caption{Ri = 2}
    \label{fig:mixing_Ri2}
  \end{subfigure}
  \begin{subfigure}{0.25\textwidth}
    \centering
    \includegraphics[width=\linewidth]{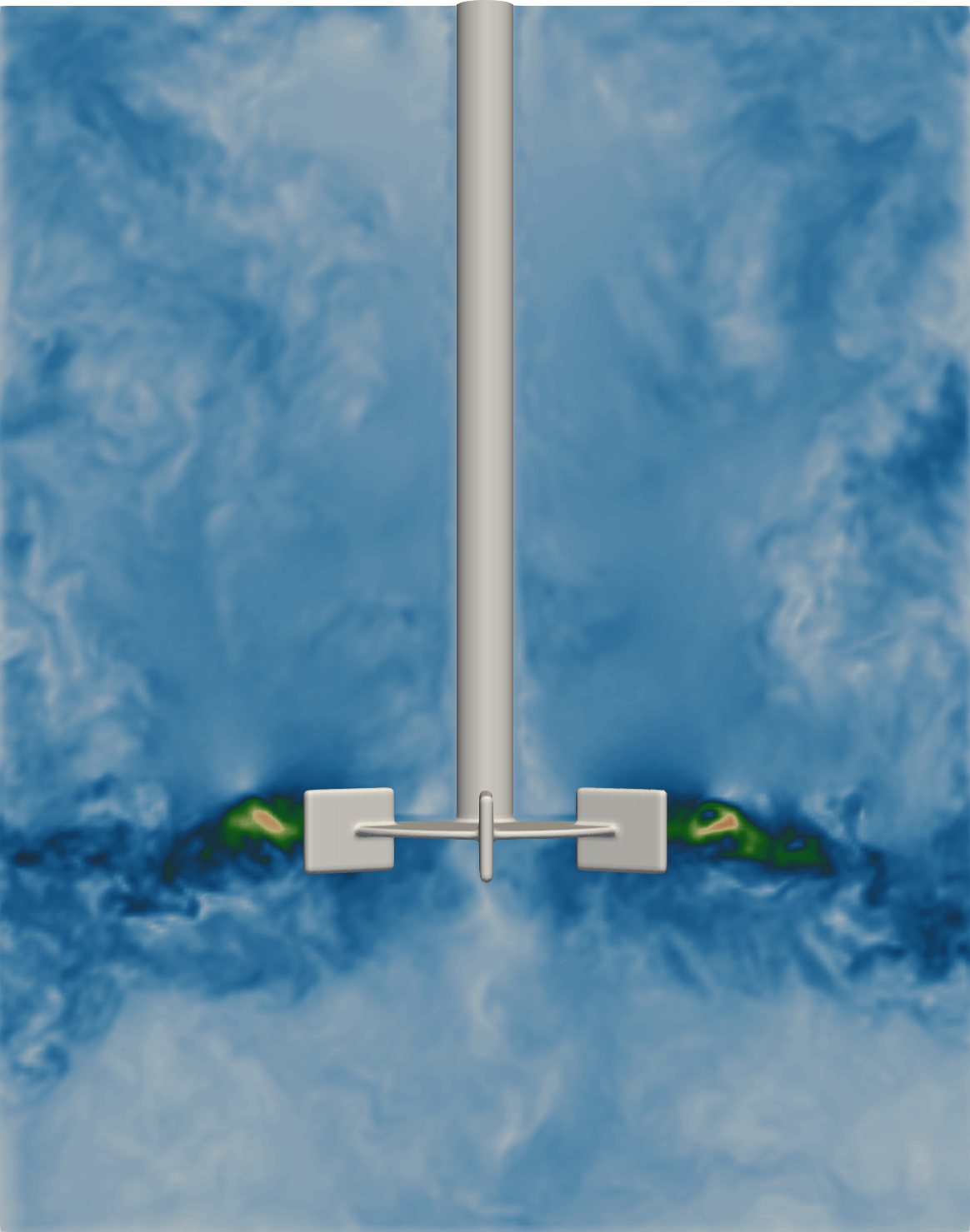}
    \caption{Ri = 6}
    \label{fig:mixing_Ri6}
  \end{subfigure}
  \begin{subfigure}{0.25\textwidth}
    \centering
    \includegraphics[width=\linewidth]{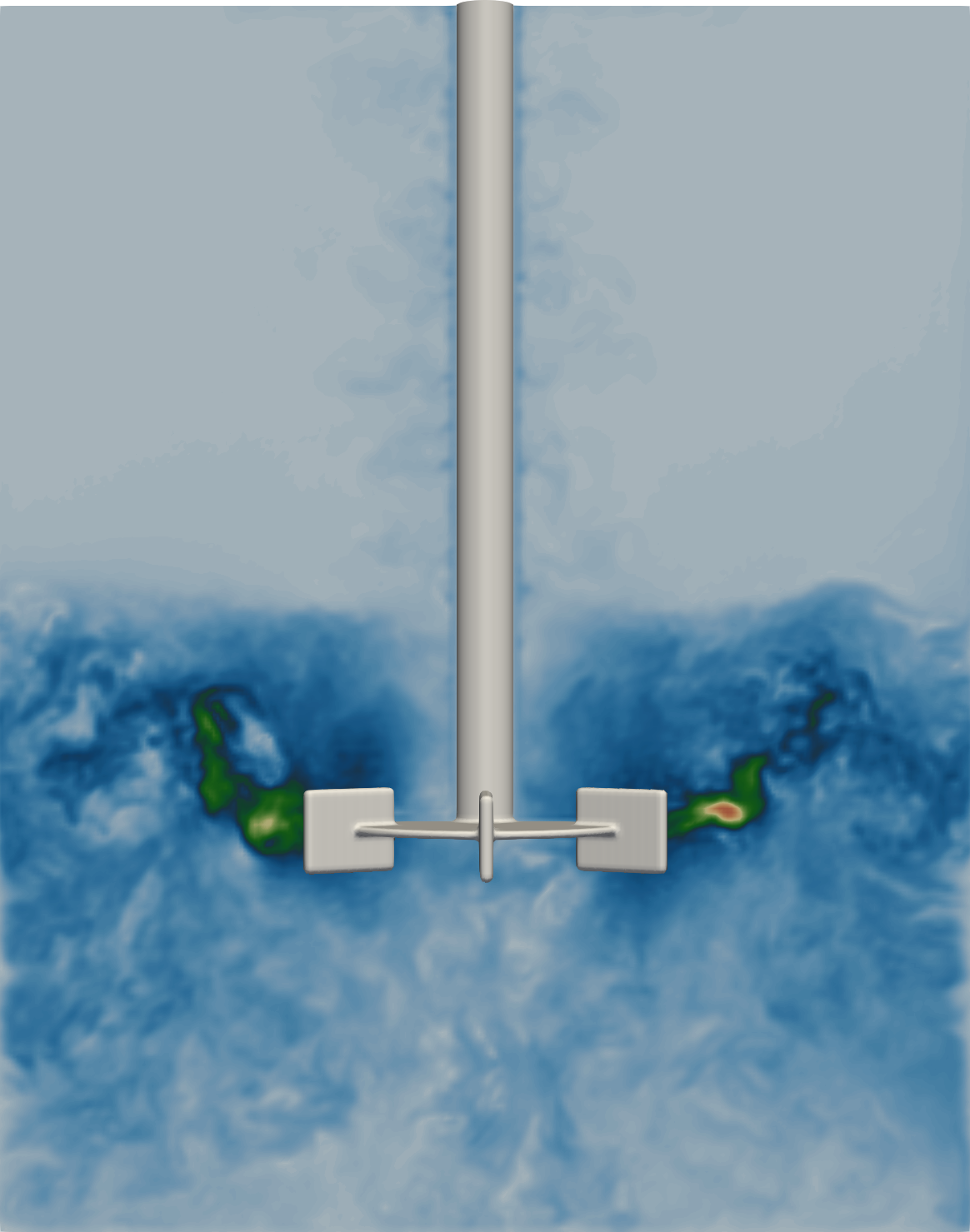}
    \caption{Ri = 12}
    \label{fig:mixing_Ri12}
  \end{subfigure}
  \begin{subfigure}{0.23\textwidth}
    \centering
    \includegraphics[height=4.8cm]{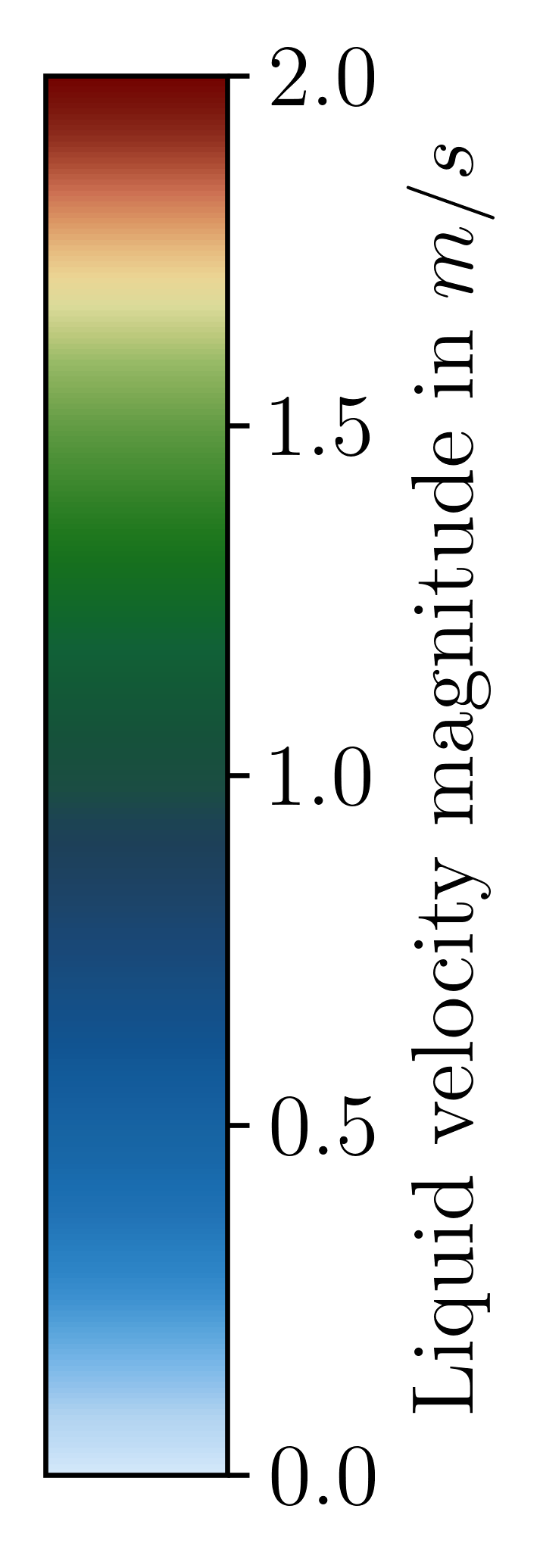}
    \vspace{8mm} 
  \end{subfigure}
  \caption{Comparison of reactor velocity cross-sections at $t = 20s$ for three Richardson numbers. Panels show (a) $Ri = 2$, (b) $Ri = 6$, and (c) $Ri = 12$, highlighting changes in flow structure with increasing buoyancy influence.}
  \label{fig:velocity_snapshots}
\end{figure}

\begin{figure}
    \centering
    \includegraphics[width=0.65\linewidth]{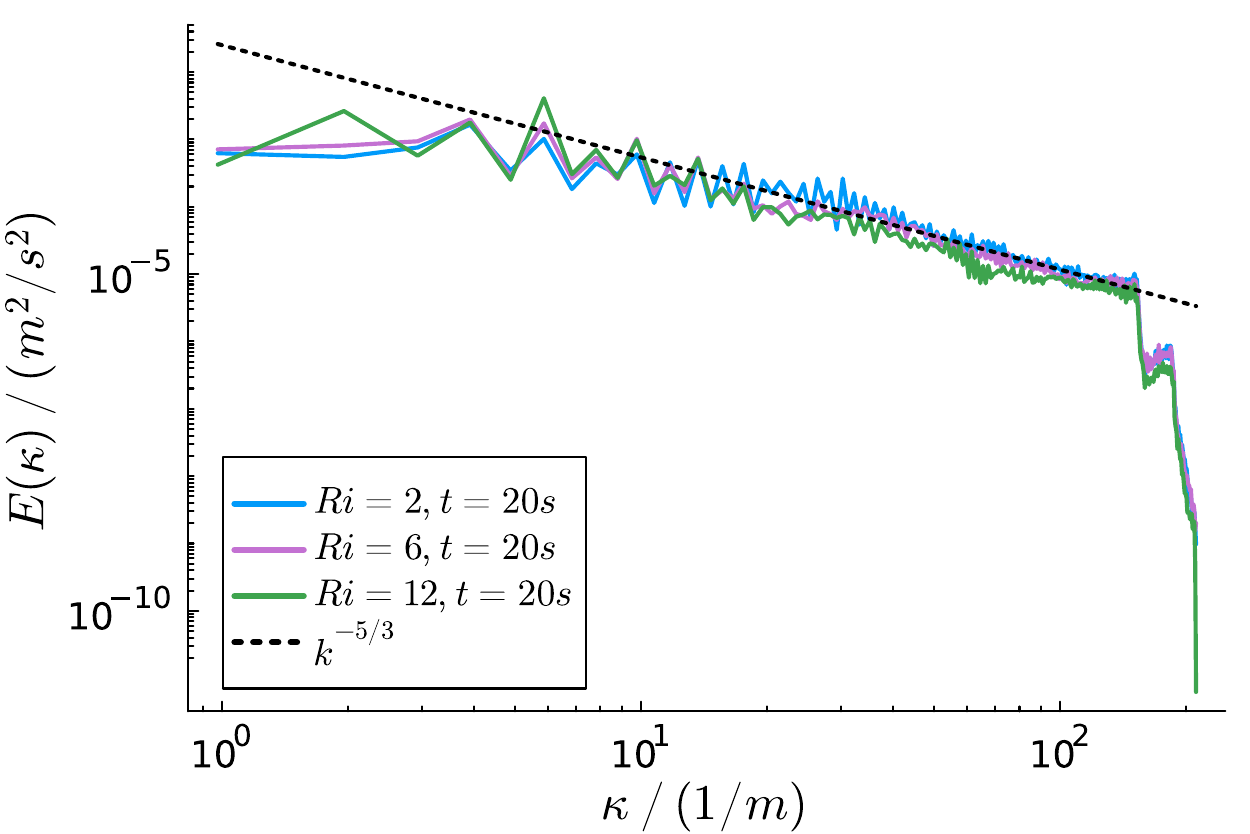}
    \caption{Power spectral analysis of the velocity field for the data shown in Fig.~\ref{fig:velocity_snapshots}(a), (b) and (c). With $\kappa$ being in the range of $[1,300]$ where 1 is set by size of the stirred tank reactor ($1m$) and 300 being the lattice resolution $0.0033m$. The spectrum starts to decline around $\kappa = 150~(0.0066m)$, well within the resolution of our system. 
    }
    \label{fig:PSD_Ri}
\end{figure}

We further quantified this by computing the power spectral density (PSD) of the velocity field (Fig.~\ref{fig:PSD_Ri}).
The limits of the PSD are on the lower end the overall size of the system, which is 1m, while on the upper end we are limited by the lattice resolution which is set to 0.0033m. 
The spectrum starts to decline well within this range at around 0.0066m for all three Richardson numbers.
Despite an increasing $Ri$, the turbulent energy spectra collapsed onto a similar inertial-range scaling, indicating that the impeller-driven dynamics dominates the energy cascade while the buoyancy primarily affects the spatial organization rather than the spectral distribution of turbulence.
The details of the PSD computation are provided in the Appendix.

\section{Conclusion}
In this work numerical simulations of turbulent mixing using the LBM were performed for the baffled stirred tank depicted in Fig.~\ref{fig:system}. 
The study covered Reynolds numbers between $5\times 10^4$ and $1.25\times 10^5$ and Richardson numbers between 0 and 14.
In the absence of buoyancy effects ($Ri=0$), the dimensionless mixing time for a homogeneity of $90\%$ was found to be $t_m^{\ast}=20$, indicating Reynolds-independent turbulent mixing in the investigated regime. 
Converting the used $90\%$ homogeneity criterion to a $95\%$ definition yields $t_{m,95}^{\ast}=26$, in agreement with literature data for turbulent mixing in baffled stirred tanks~\cite{bouwmans_blending_1997}.

For stratified systems ($Ri>0$), buoyancy significantly increases both dimensionless and dimensional mixing times. 
The simulation results collapse onto a single exponential scaling relation based on the combined influence of $N_p$, $Fr$, and $Ri$, providing a compact representation of buoyancy effect within the investigated parameter space. 
The resulting correlation enables predictive estimation of turbulent mixing times in stratified systems for the studied stirred tank configuration. 
At elevated Richardson numbers, a non-monotonic dependence of mixing time on Reynolds number was observed, indicating that increasing impeller speed does not necessarily improve mixing performance for constant Richardson numbers.

While the correlation may appear abstract it provides a simple analytical expression for engineers and application experts. 
To the best of the authors knowledge, no comparable correlation for stratified systems in the fully turbulent regime has been reported. 
The proposed correlation can help streamline mixing processes and reduce the amount of expensive real-world testing before a mixing time is defined. 
In the future we plan to test this correlation for other mixing tank geometries and include different impellers. 
Furthermore, angled stirrers create a different velocity field, it will be interesting to evaluate to what extent the proposed correlation holds for such setups as well.

\section*{Appendix}
\subsection{Grid Convergence Index}
\begin{figure}
  \centering
  \includegraphics[width=0.65\textwidth]{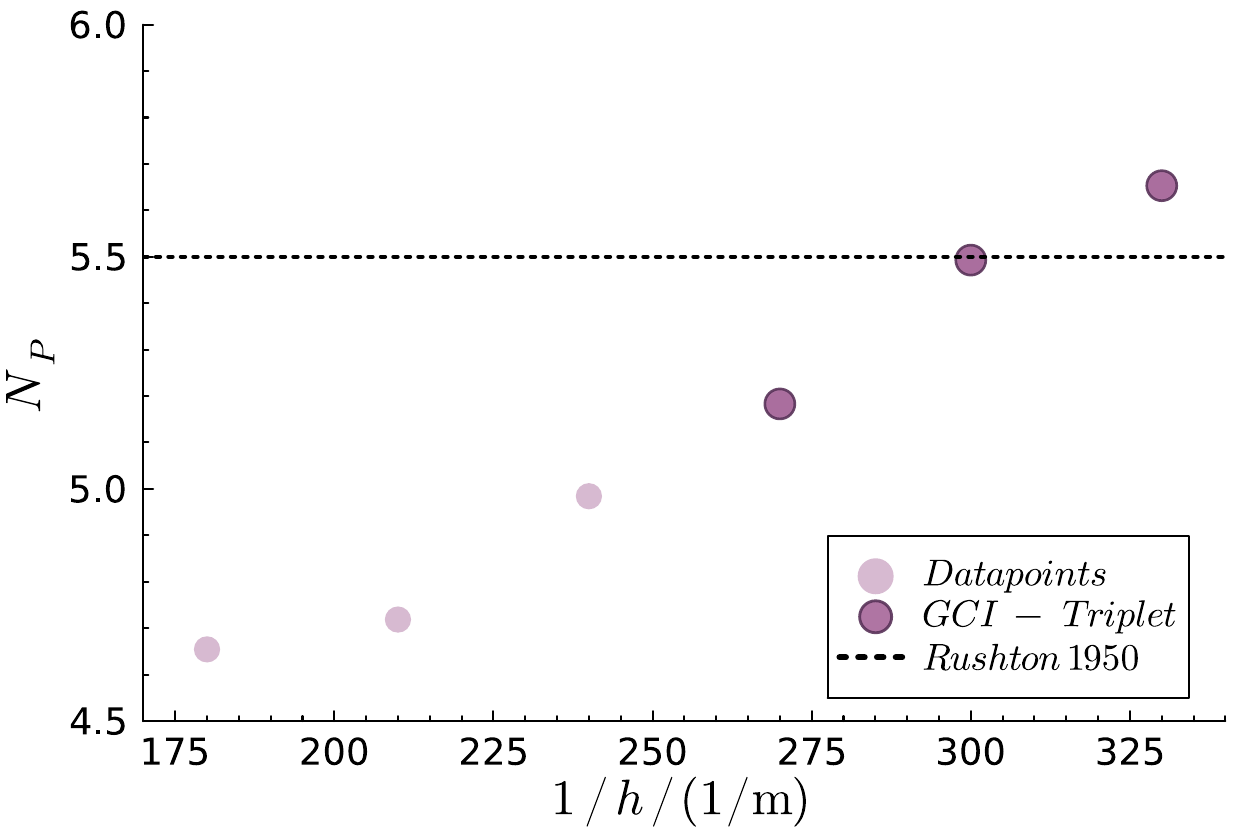}
  \caption{Power number $N_p$ versus spatial resolution $1/h$ (lattice nodes per meter). Based on a three-level GCI analysis following ASME~V\&V20, the numerical uncertainty on the production grid was estimated as $u_{\mathrm{num},21}=2.39\%$. The dashed line represents the reference power number for a Rushton turbine reported by Rushton et al.~\cite{rushtonPowerCharacteristics1950}.}
  \label{fig:power_plot}
\end{figure}

The methodology for the grid refinement study follows the procedure described in \cite{noauthor_procedure_2008}. 
Accordingly, we quantified the numerical reliability of our simulation results using a three-level grid-refinement approach and evaluated the grid-induced numerical uncertainty by means of the Grid Convergence Index (GCI). 
The quantity of interest (Solution) used for the analysis was the power number $N_p$. 
Three simulations with systematically refined grid resolutions were conducted, characterized by the grid spacing $h_1 < h_2 < h_3$. 
The associated refinement ratios were
\begin{equation}
r_{21} = \frac{h_2}{h_1}, \qquad  
r_{32} = \frac{h_3}{h_2}.
\end{equation}
The solution differences between successive grids were computed as
\begin{equation}
\varepsilon_{32} = S_3 - S_2, \qquad  
\varepsilon_{21} = S_2 - S_1.
\end{equation}
Based on these values, the apparent order of convergence $p_O$ was determined by iteratively solving the implicit Richardson extrapolation equation proposed by Roache \cite{roache_verification_nodate}.
\begin{equation}
p_O = 
\frac{1}{\ln(r_{21})}
\left|
\ln\left| \frac{\varepsilon_{32}}{\varepsilon_{21}} \right| + q(p)
\right|,
\end{equation}
with
\begin{equation}
q(p) = 
\ln\left( \frac{r_{21}^p - s}{r_{32}^p - s} \right), 
\qquad
s = \mathrm{sign}\!\left(\frac{\varepsilon_{32}}{\varepsilon_{21}}\right).
\end{equation}
The nonlinear equation was solved using a fixed-point iteration initialized with $q(p)=0$. Using the obtained convergence order, we computed the Richardson-extrapolated value of the power uptake,
\begin{equation}
S_{\text{ext}}^{21} = 
\frac{r_{21}^{p} S_1 - S_2}{r_{21}^{p} - 1},
\end{equation}
as well as the approximate relative discretization errors
\begin{equation}
e_{a,21} = \left| \frac{S_1 - S_2}{S_1} \right|,
\qquad
e_{a,32} = \left| \frac{S_2 - S_3}{S_2} \right|,
\end{equation}
and the extrapolated relative error
\begin{equation}
e_{\text{ext}}^{21}
= \left| \frac{S_{\text{ext}}^{21} - S_1}{S_{\text{ext}}^{21}} \right|.
\end{equation}
Following the standard procedure with a safety factor $F_s = 1.25$, the Grid Convergence Index on the finest grid was computed as
\begin{equation}
\mathrm{GCI}_{21}
  = \frac{F_s \, e_{21}}{r_{21}^{p} - 1}.
\end{equation}
The numerical uncertainty associated with each grid was finally estimated as
\begin{equation}
u_{\mathrm{num}} = \frac{\mathrm{GCI}}{2}.
\end{equation}
Based on a three-level grid convergence analysis following ASME~V\&V20~\cite{noauthor_procedure_2008}, the numerical uncertainty of the computed power uptake was estimated to be
$u_{\mathrm{num,21}} = 2.39\,\%$ on the selected production grid.
Accordingly, a spatial resolution of $300\,nodes\,per\,meter$ was employed for all subsequent simulations, see Fig.~\ref{fig:power_plot}.

\subsection{Mixing Time Validation}
\begin{figure}
  \centering
  \includegraphics[width=0.65\textwidth]{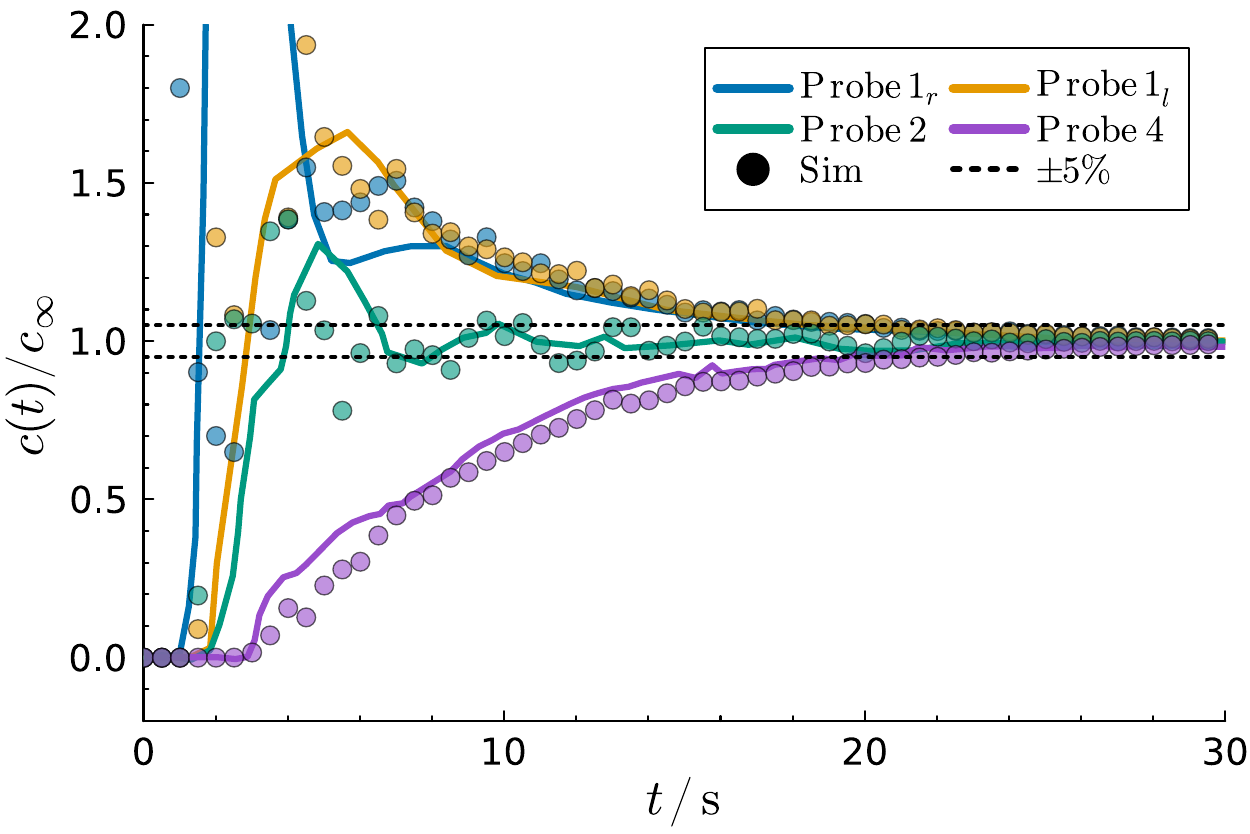}
  \caption{Mixing time validation based on probe response. 
  Experimental data (solid lines) and simulation results (symbols) are shown for different probe positions. 
  The dashed and dash-dotted horizontal lines indicate the $\pm5\%$ bounds (0.95 and 1.05) around the fully mixed state.}
  \label{fig:mixing_val_all}
\end{figure}

\begin{figure}
  \centering
  \includegraphics[width=0.65\textwidth]{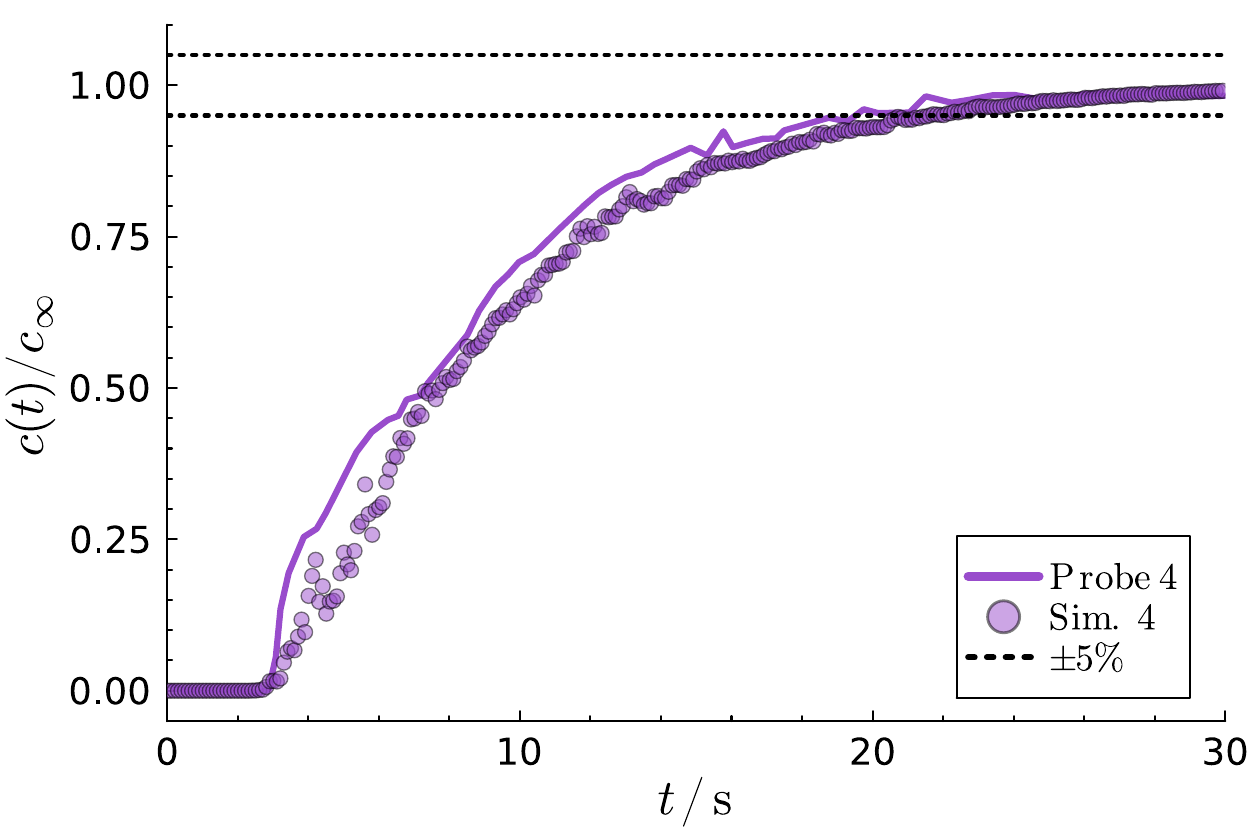}
  \caption{Mixing time validation for probe 4. 
  Experimental response (solid line) and simulation results (symbols) are compared in detail. 
  The dashed and dash-dotted lines indicate the $\pm 5\%$ bounds (0.95 and 1.05) around the fully mixed state.}
  \label{fig:mixing_val_p4}
\end{figure}

For validation purposes, the simulated concentration responses at four probe locations within the reactor are compared with experimental measurements \cite{jaworskiCFDStudyHomogenization2000}. 
The temporal evolution of the normalized probe signals demonstrates good agreement between simulations and experiments across all probe positions, capturing both the transient response and the asymptotic convergence toward the fully mixed state. 
Figure \ref{fig:mixing_val_all} summarizes the comparison between experimental and simulated probe responses for all four probe locations, illustrating the spatial variability of the mixing process within the reactor. 
Figure \ref{fig:mixing_val_p4} provides a detailed view of probe 4 as a representative example, highlighting the agreement between the simulated and experimental responses over the full transient mixing period.
\subsection*{Probe response and coefficient of variation}
To empirically determine the mixing time or the fully mixed state of a system, one can use the development of probe responses, as demonstrated by the experiments of Jaworski et al. \cite{jaworskiCFDStudyHomogenization2000}. 
For \textit{in silico} determination of such properties, the much more accurate coefficient of variation or relative standard deviation can be used. 
Investigating both quantities of the $Re~=~112500$, $Ri~=~2$ solution shows the relationship between them, as seen below. 
The relative mixing index \cite{hashmi_quantification_2014}, which is usually computed with standard deviation, is mathematically equivalent to
\begin{equation}
    RMI(t) = 1 - \frac{RSD(t)}{RSD(t=0)}.
\end{equation}
While the probe signals differ substantially during the early mixing phase, their final convergence towards homogeneity aligns with the evolution of the relative mixing index. 
In their near-homogeneous regime ($\pm10\%$ of the final value), both measures follow a similar path, and the threshold of 0.9 represents a comparable mixing state.
\begin{figure}
  \centering
  \includegraphics[width=0.65\textwidth]{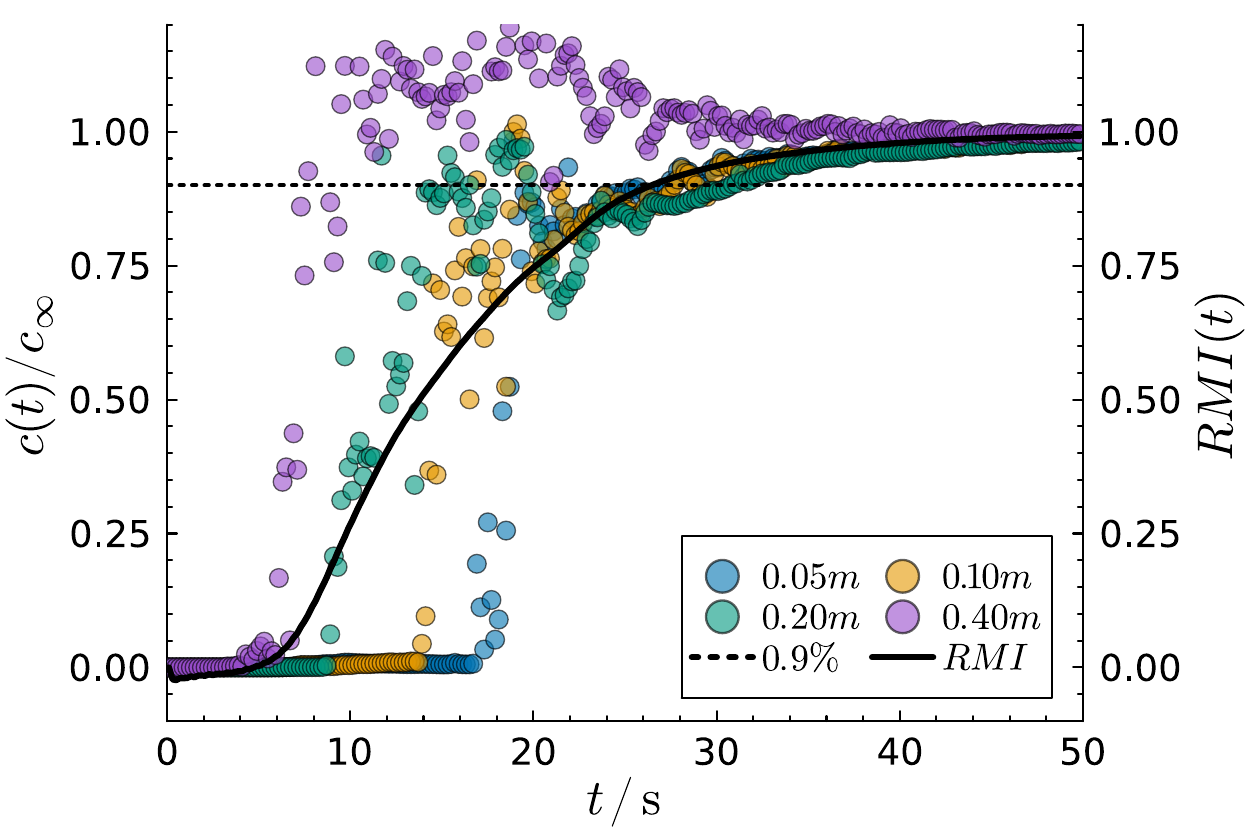}
  \caption{Comparison between the normalized probe response and the relative mixing index (RMI) for the case $Re~=~112500$, $Ri~=~2$. The probe signals were extracted at four arbitrarily selected locations positioned between two baffles at a radial distance of $r = 0.35\,\mathrm{m}$ from the reactor center and at heights of $h = 0.05\,\mathrm{m}$, $0.10\,\mathrm{m}$, $0.20\,\mathrm{m}$, and $0.40\,\mathrm{m}$. The temporal evolution of both quantities shows a close correspondence, indicating that the commonly used threshold value of 0.9 represents a comparable degree of homogenization for probe-based and RSD-based mixing measures.}
  \label{fig:mixing_index}
\end{figure}

\subsection*{Free-Surface Dynamics}
\begin{figure}
  \centering
  \includegraphics[width=0.48\textwidth]{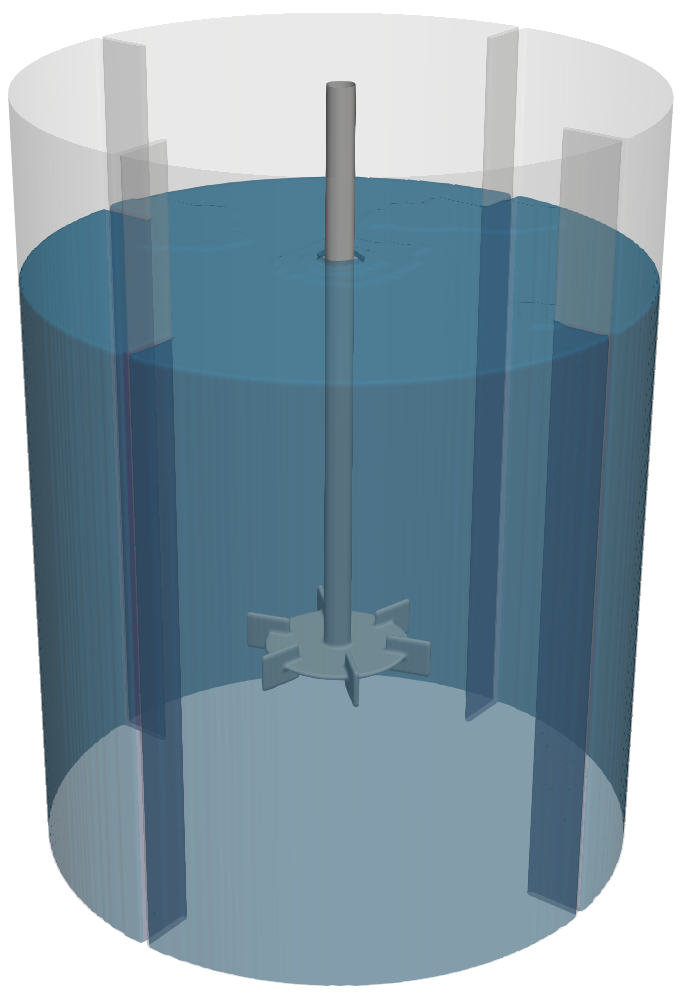}
  \caption{Instantaneous snapshot of the volume-of-fluid (VOF) simulation at $t = 19s$ for the highest Reynolds number case ($Re~=~125000$). The resolved liquid–gas interface shows only minor surface deformation, indicating a largely quiescent free surface under the investigated operating conditions.}
  \label{fig:fs_vof}
\end{figure}

To assess whether relevant free-surface effects are neglected in the single-phase simulations, the case with the highest Reynolds number was additionally investigated using a volume-of-fluid (VOF) approach.
The simulation resolves the liquid–gas interface explicitly and allows for the identification of surface deformations induced by the impeller motion. 
As shown in the figure \ref{fig:fs_vof}, the free surface remains largely flat at the investigated rotational speed, indicating that surface-related effects are negligible for this operating condition.

\subsection*{2D Spectral Analysis of a Velocity Magnitude Field}
\label{sec:psd_calc}
The computation of the energy spectra from the two-dimensional cross-section of the velocity field is described below. Since the velocity fields shown in Figure 5 was used in this computation, the coordinates and the velocity magnitude were extracted from the y-plane results for the reactor. With the structured velocity values, we performed a fft ($\hat{\mathbf{v}}(\kappa_x,\kappa_z)$). The power spectral density is computed according to:
\begin{equation}
\mathrm{PSD}(\kappa_x, \kappa_z)
= \frac{|\hat{v}(\kappa_x,\kappa_z)|^2}{N_x N_z}
\, \Delta x \, \Delta z,
\end{equation}
where the normalization is chosen such that the discrete form of Parseval's theorem is satisfied,
\begin{equation}
\sum_{x,z} |v'(x,z)|^2 \, \Delta x \, \Delta z
=
\sum_{\kappa_x,\kappa_z}
\mathrm{PSD}(\kappa_x,\kappa_z).
\end{equation}
The scalar wavenumber is defined as:
\begin{equation}
\kappa = \sqrt{\kappa_x^2 + \kappa_z^2}
\end{equation}
Spectral bins are constructed using the minimum resolvable wavenumber increment:
\begin{equation}
\Delta \kappa = \min(\Delta \kappa_x, \Delta \kappa_z)
\end{equation}
The isotropic energy spectrum is evaluated as:
\begin{equation}
E(\kappa)
= 2\pi \kappa
\frac{1}{N(\kappa)}
\sum_{\kappa_i \in \kappa}
\mathrm{PSD}(k_x,k_z)
\end{equation}
where $N(\kappa)$ denotes the number of spectral modes per bin.

\section*{CRediT authorship contribution statement}
\textbf{M. R. Wagner}: Conceptualization; Methodology; Software; Validation; Formal analysis; Investigation; Writing – Original Draft; Writing – Review \& Editing; Visualization. 
\textbf{M. Dubacher}: Conceptualization; Writing – Review \& Editing. 
\textbf{N. Patsaki}: Conceptualization; Methodology. 
\textbf{P. Eibl}: Software; Validation; Writing – Review \& Editing. 
\textbf{P. V. Dsouza}: Conceptualization; Formal analysis; Investigation; Writing – Original Draft. 
\textbf{M. Dekner}: Writing – Review \& Editing. 
\textbf{C. Witz}: Software. 
\textbf{J. Remmelgas}: Conceptualization; Writing – Review \& Editing. 
\textbf{S. Reimann-Zitz}: Conceptualization; Methodology; Formal analysis; Investigation; Writing – Original Draft; Writing – Review \& Editing; Visualization. 
\textbf{J. Khinast}: Resources; Writing – Review \& Editing; Supervision.

\section*{Acknowledgements}
The Research Center Pharmaceutical Engineering (RCPE) is funded within the framework of COMET - Competence Centers for Excellent Technologies by BMK, BMAW, Land Steiermark, and SFG. 
The COMET program is managed by the FFG.
The authors thank Nicolas Morse for fruitful discussions.

\printbibliography

\newpage
\section*{Graphical Abstract}
\begin{figure}[h]
  \centering
  \includegraphics[height=4.76cm]{Fig3_4.png}
  \hspace{0mm}%
  \includegraphics[height=4.76cm]{Fig3_2.png}
  \hspace{1mm}%
  \includegraphics[height=4.76cm]{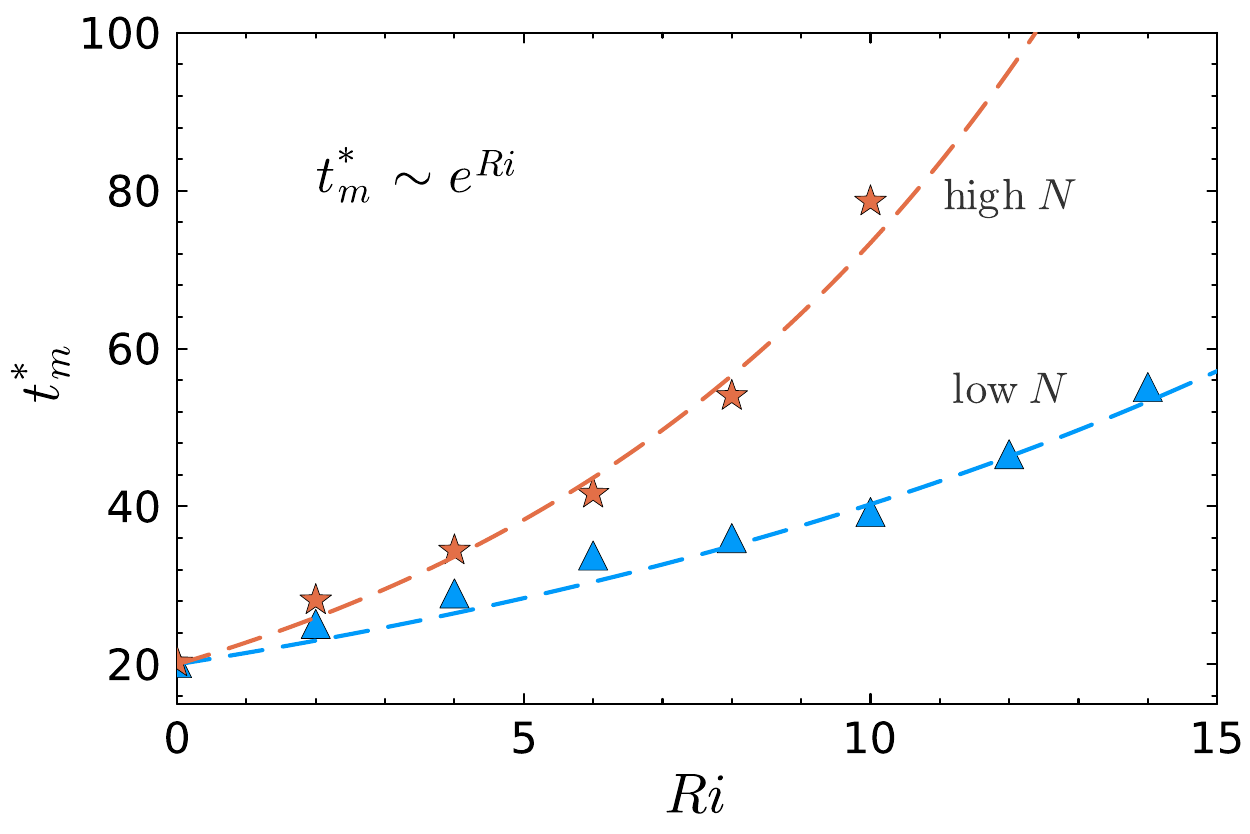}
  \label{fig:TOC_graphic}
\end{figure}

\end{document}